\documentclass{elsart}
\usepackage{epsfig}
\usepackage{color}
\usepackage{amsmath}
\usepackage{amsfonts}
\usepackage{graphicx,slashbox}
\usepackage{ulem}

\definecolor{green2}{rgb}{0,0.5,0}

\begin{document}

\begin{frontmatter}

\title{On the exact treatment of Time Dependent Self-Interaction Correction}

\author{J. Messud\corauthref{cor}$^{a,b}$}, 
\author{P.~M.~Dinh$^{a,b}$, P.-G.~Reinhard$^c$, E.~Suraud$^{a,b}$}

\corauth[cor]{Corresponding author\\{\it Email-address}~:
  messud@irsamc.ups-tlse.fr} 
\address{$^a$Universit\'e de Toulouse; UPS; \\Laboratoire de Physique
  Th\'eorique (IRSAMC); F-31062 Toulouse, France}
\address{$^b$ CNRS; LPT (IRSAMC); F-31062 Toulouse, France}
\address{$^c$Institut f{\"u}r Theoretische Physik, Universit{\"a}t
  Erlangen, D-91058 Erlangen, Germany}

\begin{abstract}
We present a new formulation of the time-dependent self-interaction
correction (TDSIC). It is derived variationally obeying explicitly
the constraints on orthonormality of the occupied single-particle
orbitals. The thus emerging rather involved symmetry condition
amongst the orbitals is dealt with using two separate sets of
(occupied) single-particle wavefunctions, related
by a unitary transformation. The double-set TDSIC scheme
is well suited for numerical implementation. We present
results for laser-excited dynamics in a 1D model for a molecule
and in fully fledged 3D calculations. 
\end{abstract}

\end{frontmatter}

%% \begin{center}
%% \begin{minipage}{9cm}
%% \PGRcomm{\it
%% In that versions, all footnotes have been removed which
%% seemed to have found agreeable answers.
%% Remaining and new footnotes appear need yet to be resolved.
%% \\
%% All previous author colors have been removed. New
%% writings are again colored appropriately.
%% \\
%% The empty entries in 'add.bib' should be filled.
%% I do not know who is in posession of the corresponding details,
%% for sure not me.
%% \\
%% Some equations exceed the line width. That needs still
%% to be aligned.
%% \\
%% Spell checking has to be done.
%% }
%% \end{minipage}
%% \end{center}

\section{Introduction}
\label{sec:intro}

Density Functional Theory (DFT)~\cite{Hoh64,Par89,Dre90,Koh99r} has
become over the last decades a widely used theoretical tool for the
description and analysis of electronic properties in physical and
chemical systems. This applies particularly to systems with
sizeable numbers of electrons \cite{Par89,Dre90}, all the more so if
one is interested in truly dynamical situations. The
extension to Time-Dependent DFT (TDDFT) has been formally established
more recently \cite{Run84,Gro90,Mar04} and it is still in
development, concerning both formal and practical aspects
\cite{Mar06}.  Over the years, TDDFT has thus become one of the few,
well founded theories, allowing to describe dynamical scenarios in
complex systems. This is a key issue for understanding
dynamical microscopic mechanisms, beyond mere energetic
considerations, e.g.  the process of electron emission as it is
important in connection with laser irradiation.

A basic idea of (TD)DFT is to replace the involved correlated
many-electron problem by an effective one-body description though the
inclusion of exchange and correlation effects in a (as simple as
possible) exchange and correlation {functional, expressed
in terms of the local density of the electrons.}  The simplest
approximation to the exchange correlation functional is the Local
Density Approximation (LDA), or Adiabatic Local Density Approximation
(ALDA) in the time dependent case, which proved very useful in
calculations of structure and low-amplitude excitations (optical
response, direct one-photon processes) \cite{Koh99r}.  It can also be
used as a first order approach in more violent dynamical processes
involving huge energy deposits and/or large ionization as, for
example, in the case of clusters or molecules subject to intense laser
fields or to collision with highly charged particles \cite{Rei03a}.

However, LDA is plagued by a self-interaction error due to the fact
that the direct Coulomb term and the exchange-correlation potential
involve the total density including the particle on which the field
actually acts. 
{ That Coulomb self-interaction is nicely canceled in a full
Hartree-Fock treatment. However, the approximate treatment of exchange
in LDA weakens this cancellation and a spurious self-interaction
remains.
}
As a consequence, LDA produces the wrong Coulomb asymptotic.  The
self-interaction thus spoils single-particle properties as, e.g., the
Ionization Potential (IP) or the band gap in solids
\cite{Hyb86a,Nie00aR}. Another critical detail where LDA fails is the
polarizability in chain molecules \cite{Gis99a,Kue04a}.  In dynamical
situations, the self-interaction error will {thus} spoil the
description of excitations involving ionization processes, especially
in processes close to electron emission threshold.  Correcting the
self-interaction error requires a dedicated treatment known as the
Self-Interaction Correction (SIC). Such a SIC complementing LDA static
calculations was proposed in \cite{Per79a,Per81}. It has been used
since then at various levels of refinement for structure calculations
in atomic, molecular, cluster and solid state physics, see
e.g. \cite{Ped84,Goe97,Polo,Vyd04}. The original SIC scheme, however,
leads to an orbital dependent mean field which causes several formal
and technical difficulties. This aspect can be circumvented by
treating SIC with optimized effective potentials (OEP), 
see~\cite{Kue07} for a recent review. The resulting formalism
is quite involved and usually treated with involving further
approximations, as e.g. the Krieger-Li-Iafrate (KLI)
approach~\cite{Kri92a,Kri92b}. These,  
however, can perturb some crucial physical features of SIC,
particularly the trend to produce localized single-particle states
\cite{Kue07}.

The applications of SIC in time-dependent situations have, up to now,
mostly been performed in the above mentioned approximate manners,
e.g., the {linearized} treatment of \cite{Pac92}, the use of
averaged-density SIC \cite{Leg02}, or the various versions of time
dependent OEP-KLI \cite{Ull95a,Ton97,Ton01}. The latter TDOEP-KLI,
however, also suffers from inconsistencies. It leads, in particular,
to the violation of zero force theorem and energy conservation
\cite{Mun07a}.  There thus remains plenty of space for elaborating
adequate versions of SIC, {and even more so of TDSIC.  The
natural starting point, and benchmark for later approximations is a
full TDSIC scheme.}  The aim of this paper is to present an exact
thorough variational formulation of {fully fledged} TDSIC.  We shall
also propose a {manageable} propagation scheme which allows to obey
all key conditions, {namely the zero-force theorem, conservation of
energy and conservation of orthonormality of the occupied
single-particle orbitals.} Our approach relies on a simple account of
basic constraints and on the use of an important degree of freedom,
namely the freedom of unitary transforms among occupied orbitals.

The paper is organized as follows. We first remind
basic SIC equations in static case, introducing already the unitary
transform degree of freedom. We then export the formalism in the time
domain and discuss the properties of TDSIC. We finally show practical
examples of applications in simple molecules and clusters, in
particular in the case of irradiation processes.

\section{Basic notations}
\label{sec:basics}

Before attacking the question of the self-interaction correction, we
want to introduce briefly generic notations and take the example of
widely used Local Density Approximation (LDA) \cite{Jon89},
\cite{Koh99} which serves as a basis for our further considerations.
We shall work in the Kohn-Sham scheme of DFT \cite{ks}. 
The Kohn-Sham state is composed of a set single-particle
wavefunctions $\{\psi_\alpha, \, \alpha=1,\ldots,N\}$ where $N$ is the
number of electrons of the system.  These single-particle states have
to be orthonormalized. This requirement will play a role later on.
Both static and dynamical DFT schemes then amount to write effective
one-body Schr\"odinger-like equations for the $\{\psi_\alpha, \,
\alpha=1,\ldots,N\}$, called Kohn-Sham (KS) equations.  In the KS
scheme, the total electronic energy of the system $E$ can be split
into four terms :
\begin{equation}
E = E_{\rm kin} + E_{\rm ext} + E_{\rm H} + E_{\rm xc}.
\end{equation}
The kinetic component $E_{\rm kin}$ is computed assuming non
interacting $\psi_\alpha$; the direct Coulomb interaction $E_{\rm H}$
is computed computed classically (Hartree approximation
\cite{hartree}); the effect of the external potential ($E_{\rm ext}$
including in particular the ionic potential and possibly external
fields such as that delivered by a laser) is computed exactly and
finally the exchange Coulomb and the electronic correlations are
packed into the exchange correlation energy $E_{\rm xc}$ for which one
has to construct approximations.  All terms are functionals of the
total electronic density $\rho({\bf r})$.

The simplest and most widely used approximation for $E_{\rm xc}$ is
the Local Density Approximation (LDA) in which one performs a local
Fermi gas approximation for evaluating energies. The LDA serves as a
starting point for many more involved approximations, in particular
the SIC approximation we discuss in this paper. We thus assume that
$E_{\rm xc}$ is computed in the LDA approximation. For the sake of
simplicity in the notations, we shall pack together the (exact) Hartree
and exchange correlation terms and note $E_{\rm LDA}$ the
corresponding energy at LDA approximation : $E_{\rm LDA}=E_{\rm H} +
E_{\rm xc}$. One can then derive the KS equations (stationary or time
dependent) by standard variational techniques which leads to single
electron KS equations with LDA single electron Hamiltonian
\begin{eqnarray}
  \hat{h}_\mathrm{LDA}
  &=&
  -\frac{\hbar^2 \Delta}{2m}
  + U_\mathrm{ext} + U_\mathrm{LDA}[\rho]
  \quad,
\\
  U_\mathrm{LDA}[\varrho]
  &=&
  \left.\frac{\delta E_\mathrm{LDA}}{\delta\rho}\right|_{\rho=\varrho} 
  = U_{\rm H}[\varrho]+U_{xc}[\varrho]
  \quad.
\end{eqnarray}
{As outlined in the introduction, the LDA approximation
suffers from the self-interaction error.  For example, its direct
Coulomb part $U_{\rm H}[\rho]$ is a functional of the total density
which is 
computed by summing over all occupied single electron densities ($\rho
= \sum_\alpha \rho_\alpha$).  Then a particle $\alpha$ will feel its
own Coulomb repulsion and this spurious self-interaction is not
properly removed by the exchange term in LDA.  This thus calls for a
SIC treatment.}
 
\section{Stationary SIC}
\label{sec:sic}

\subsection{SIC functional and Hamiltonian}
\label{sec:sic_notations}

The starting point is the SIC
energy functional (following notations of section \ref{sec:basics})
\begin{eqnarray}
  E_\mathrm{SIC}
  =
  E_\mathrm{kin}+E_\mathrm{ext}
  \!+\!
  E_\mathrm{ion}
  \!+\!
  E_\mathrm{LDA}[\rho]
 %\nonumber \\
  \!-\!
  {\sum_{\beta=1}^{N} E_\mathrm{LDA}[|\psi_{\beta}|^2]}
\label{eq:fsicen}
\end{eqnarray}
{where the total density is $\rho=\sum_{\alpha} \rho_{\alpha}$
with $\rho_\alpha=|\psi_{\alpha}|^2$}.  Note that all summations run
over occupied states only.
%\section{Stationary SIC}
The corresponding one-body  Hamiltonian is obtained from variation of
$E_\mathrm{SIC}$ with respect to $\psi^*_{\alpha}$ as
\begin{subequations}
\label{eq:mfham}
\begin{eqnarray}
  \frac{\delta E_\mathrm{SIC}}{\delta\psi_\alpha^*}
  &=&
  \hat{h}_\alpha\psi_\alpha
  \;,\;\\
  \hat{h}_\alpha
   &=&
  \hat{h}_\mathrm{LDA}-U_\alpha
  \;,
\label{eq:halpha}\\
  U_\alpha 
  &=&  
  U_\mathrm{LDA}[|\psi_{\alpha}|^2]
  \quad.
\label{eq:usic}
\end{eqnarray}
\end{subequations}
The emerging one-body Hamiltonian $\hat{h}_\alpha$ depends on the
state $\psi_\alpha$ on which it acts through the SIC term $U_\alpha$.
Thus it is not invariant {under} unitary transformations
{within the sub-space of occupied orbitals}.

\subsection{{The stationary SIC equations}}
\label{sec:sic_1set}

\subsubsection{Variational derivation}
\label{sec:vary-single}

The static SIC equations are derived by minimization of the
SIC energy (\ref{eq:fsicen}) together with the condition that
the single-particle orbitals are orthonormalized.
This amounts to the variational equation
\begin{equation}
  0
  =
  \delta_{\psi_\alpha^*}
  \left[
   E_\mathrm{SIC} 
   - 
   \sum_{\alpha,\beta}(\psi_\alpha|\psi_\beta)\lambda_{\beta\alpha}
  \right]
\end{equation}
where $\lambda_{\alpha\beta}$ is a matrix of Lagrangian multipliers,
which is non-diagonal in general.  As worked out in appendix
\ref{app:hermconstr}, $\lambda_{\alpha\beta}$ is a hermitian matrix.
Evaluation of the variation yields the stationary equations
\begin{subequations}
\label{eq:SICeqs}
\begin{eqnarray}
  \hat{h}_\alpha|\psi_\alpha)
  &=&
  \sum_{\beta} |\psi_{\beta})
  \lambda_{\beta\alpha} 
  \quad,
\label{eq:statSIC1}\\
  \lambda_{\beta\alpha}
  &=&
  (\psi_\beta|\hat{h}_\alpha|\psi_\alpha) 
  \quad.
\end{eqnarray}
\end{subequations}
We consider the hermitian conjugate equation
$$
  (\psi_\beta|\hat{h}_\beta
  =
  \sum_\alpha\lambda_{\alpha\beta}^*(\psi_\alpha|
  =
  \sum_\alpha\lambda_{\beta\alpha}(\psi_\alpha|
$$ 
(where we have exploited hermiticity of $\lambda_{\beta\alpha}$),
project Eq. (\ref{eq:statSIC1}) with $(\psi_\beta|$, its conjugate
with $|\psi_\alpha)$, and take the difference of these two equations.
This yields
$
  0
  =
  (\psi_\beta|\hat{h}_\beta-\hat{h}_\alpha|\psi_\alpha)
  \,.
$
The only state-dependence in $\hat{h}_\alpha$ stems from
$U_\alpha$. Thus we remain with the condition
\begin{equation}
  0
  =
  (\psi_\beta|U_\beta-U_\alpha|\psi_\alpha) 
  \quad.
\label{eq:symcond1}
\end{equation}
We call it the {\em symmetry condition}. It plays a crucial role in
all SIC considerations.
It was first introduced in a particular case by Pederson {\it et al.}
\cite{Ped84} and since then addressed by several authors
\cite{Goe97,Vyd04,Jon89}.
The above derivation indicates clearly the relation between
symmetry condition Eq. (\ref{eq:symcond1}) and orthonormality
constraint.  We shall discuss this condition further at
several places.
Note that it is trivially fulfilled in case of state-independent
Hamiltonians for which $\hat{h}_\alpha=\hat{h}_\beta=\hat{h}$, as it is
the case in LDA or Hartree-Fock.

A word is in order about the SIC Hamiltonian $\hat{h}_\alpha$.  It
depends on the state on which it acts. Thus one has to be extremely
careful with everything one knows from Quantum Mechanics and Hilbert
space. The $\hat{h}_\alpha$ is not a linear operator which is obvious
from the fact that the operation
$\hat{h}_\alpha \left[|\psi_\alpha)c_\alpha+|\psi_\beta)c_\beta
  \right]$ 
is not defined at all. We will also see more clearly in the next
section that the SIC Hamiltonian is not hermitian.

\subsubsection{State-independent notation}
\label{sec:indepSIC}

For formal manipulations, it may be simpler to recast the
{state dependent} SIC Hamiltonian $\hat{h}_\alpha$ into a
compact form as
\begin{eqnarray}
  \hat{h}_\mathrm{SIC}
  &=&
  \hat{h}_\mathrm{LDA}
  -
  \sum_\alpha U_\alpha|\psi_\alpha)(\psi_\alpha| 
  \quad.
\label{eq:hsic}
\end{eqnarray}
The part sensitive to single-particle states has been expressed in
terms of projectors $|\psi_\alpha)(\psi_\alpha|$ such that the SIC
Hamiltonian (\ref{eq:hsic}) is not explicitly state-dependent, but
$\hat{h}_\mathrm{SIC}|\psi_\alpha)$ remains equivalent to
$\hat{h}_\alpha|\psi_\alpha)$. The form (\ref{eq:hsic}) is advantageous
for formal considerations.
The SIC equations become now equivalently
\begin{subequations}
\label{eq:statsic1}
\begin{eqnarray}
  \hat{h}_\mathrm{SIC}|\psi_\alpha)
  &=&
  \sum_{\beta} |\psi_{\beta})
  \lambda_{\beta\alpha} 
  \ ,
\label{eq:statsic}\\
  \lambda_{\beta\alpha}
  &=&
  (\psi_\beta|\hat{h}_\mathrm{SIC}|\psi_\alpha) 
  \ .
\label{eq:constrmat}
\end{eqnarray}
The symmetry condition is derived as above.  We build the
hermitian conjugate equation, take the difference of the two
equations, and exploit the fact that the Lagrangian matrix is
hermitian. This yields
\begin{equation}
 0
 =
 (\psi_\beta|\hat{h}_\mathrm{SIC}^\dagger
 -
  \hat{h}_\mathrm{SIC}|\psi_\alpha)
\label{eq:symcond2}
\end{equation}
\end{subequations}
which is equivalent to the symmetry condition in the form
(\ref{eq:symcond1}) when dropping the state independent part
$\hat{h}_\mathrm{LDA}$ and evaluating the projectors.

It is interesting to note that the form (\ref{eq:hsic}) shows
clearly the possible non-hermiticity of the SIC Hamiltonian.
This makes the symmetry condition (\ref{eq:symcond2}), or
equivalently (\ref{eq:symcond1}), a non-trivial and crucial part 
of the SIC equations.
{Superficially, it makes the impression of a condition
ensuring hermiticity of the SIC Hamiltonian $\hat{h}_\mathrm{SIC}$.
But one has to keep in mind that condition (\ref{eq:symcond2}) is
restricted to the space of occupied orbitals. Thus
the symmetry condition forces
restoration of hermiticity only in the sub-space of occupied
states.

\subsubsection{Projector notation}
\label{sec:proj}

The SIC equation can be recast in a particularly compact form when
introducing the projection operator onto the unoccupied space
\begin{equation}
  \hat{\Pi}_\perp 
  =  
  \hat{1}-\sum_\beta|\psi_\beta)(\psi_\beta|
  \ .
\label{eq:proj}
\end{equation}
{
This allows to reformulate the SIC equations (\ref{eq:statSIC1})
or (\ref{eq:statsic}) as
}
\begin{equation}
  \hat{\Pi}_\perp \hat{h}_\alpha |\psi_\alpha) 
  = 
  0 
  \qquad {\rm or}\qquad
  \hat{\Pi}_\perp   \hat{h}_\mathrm{SIC}|\psi_\alpha) 
  = 
  0 
\label{eq:statsic3}
\end{equation}
showing that these equations serve to establish a decoupling
of occupied and unoccupied space which is a general feature of any
mean-field equation.  The new key feature of the SIC equations is the
additional symmetry condition, Eqs. (\ref{eq:symcond1}) or
(\ref{eq:symcond2}), which comes into play because the SIC energy
(\ref{eq:fsicen}) is not unitary invariant such that there is a unique
optimum for the occupied states.

\subsubsection{Single-particle energies}
\label{sec:spe}

The symmetry condition minimizes the SIC energy and does that by
producing more or less localized states which maximize the Coulomb SIC
of each state (see the later discussion).  This produces in general
non-diagonal Lagrangian matrices $\lambda_{\alpha\beta}$ from which
single-particle energies cannot immediately be read off. However, the
necessary information is contained in that matrix.  The
single-particle energies can be defined as the eigenvalues
$\varepsilon_i$ of $\lambda_{\alpha\beta}$ obtained from the secular
equation in occupied space
\begin{equation}
  \sum_\beta\lambda_{\alpha\beta}v_{\beta i}
  =
  \varepsilon_i v_{\alpha i}
\end{equation}
where $v_{\alpha i}$ are coefficients of the appropriate unitary
transformation.

\subsection{Double-set formulation of SIC}
\label{sec:sic_2set}

Thus far, the formulation of SIC for stationary states is
complete and manageable. The computation of single-particle energies
motivates an alternative formulation which deals with two different,
but related, sets of occupied single-particle states. We will thus
discuss in this section a double-set formulation of stationary SIC.
It is an interesting, but not compulsory, alternative for the static
case. But a double-set technique becomes almost inevitable for TDSIC.
The present (static) section serves, so to say, as a preparation.

\subsubsection{Two sets of occupied states}

The computation of single-particle energies, as outlined in section
\ref{sec:spe}, leads naturally to a second set of single-particle
states $\{\varphi_i\}$ connected to the original set by a
unitary transformation within occupied space
\label{eq:statsic2}
\begin{equation}
  \varphi_i
  =
  \sum_{\alpha=1}^N\psi_\alpha v_{\alpha i}
  \quad,\quad
  \sum_\alpha v_{\alpha i}^*v_{\alpha j}=\delta_{ij}
  \ .
\label{eq:unitrans}
\end{equation}
The set $\{\varphi_i\}$ is associated to the single-particle energies
$\varepsilon_i$ and so diagonal in energy space. We call it {\em
diagonalizing set}. The set $\{\psi_\alpha\}$ optimizes the SIC
potentials and does that by some localization. We call it the {\em
localizing set}. The diagonalizing set is compact in energy space at
the price of larger spatial spreading and the localizing set minimizes
spatial extension while enhancing energy variance.  Both sets have
their value. A proper combination of them will become particularly
important in the dynamical case, see section \ref{sec:TDSIC}.

\subsubsection{Double-set SIC equations}
\label{sec:dsetSIC}

The first SIC equation (\ref{eq:statsic}) becomes particularly
simple in terms of the diagonalizing set. It reads now
\begin{subequations}
\label{eq:doubleseteqs}
\begin{equation}
  \hat{h}_\mathrm{SIC}|\varphi_i)
  =
  \varepsilon_i |\varphi_i)
  \quad.
\label{eq:static-diaq}
\end{equation}
That equations provides the decoupling from unoccupied space as shown
in section \ref{sec:proj}.  The symmetry condition cares for
determining the localizing set within occupied space which now shrinks
to a condition for the transformation coefficients $v_{\alpha i}$. We
emphasize that by rewriting
\begin{equation}
  v_{i\alpha}
  \quad\longleftrightarrow\quad
  0
  =
  (\psi_\beta|U_\beta-U_\alpha|\psi_\alpha) 
  \ .
\label{eq:symcond2b}
\end{equation}
\end{subequations}
That localizing set is needed to compute the SIC potentials
$U_\alpha$ and with it, $\hat{h}_\mathrm{SIC}$.

The double-set equations (\ref{eq:doubleseteqs}) can be used for an
alternative solution scheme. However, there is no gain in efficiency
as compared to the previous scheme, i.e. solving first the SIC
equations (\ref{eq:SICeqs}) with a single set $\{\psi_\alpha\}$ and
afterwards diagonalizing the Lagrangian matrix $\lambda_{\beta\alpha}$
to obtain the single-particle energies.

\subsubsection{Variational derivation}

It is instructive to derive double-set SIC directly from the
stationary variational principle. To that end, we consider the
diagonalizing set $\{\varphi_i\}$ and the transformation coefficients
$v_{\alpha i}$ as variational degrees of freedom. The SIC functional
is to be minimized with boundary conditions of orthonormality of the
$\varphi_i$ and $v_{\alpha i}$. This means to minimize the functional
\begin{equation}
  F[\varphi_i,v_{\alpha i}]
  =
  E_\mathrm{SIC}[\varphi_i,v_{\alpha i}]
  -
  \sum_{k,j}(\varphi_k|\varphi_j)\theta_{jk}
  -
  \sum_{\alpha\beta}
    \big(\sum_i v_{\alpha i}^*v_{\beta i}^{\mbox{}}\big)
    \Lambda_{\beta\alpha}
 \ .
\label{eq:varprinconstr-dsic-static}
\end{equation}
It is interesting to compare the extension with LDA.  In that case, one
deals with an energy functional which is invariant under
unitary transformations amongst occupied states. That allowed to
perform always a unitary transformation such that
$\sum_{k,j}(\varphi_k|\varphi_j)\theta_{jk}
 \longrightarrow
 \sum_{j}(\varphi_j|\varphi_j)\varepsilon_{j}
$
from which one obtains immediately the energy-diagonal LDA equations
by variation.  The SIC functional (\ref{eq:fsicen}) is not unitary
invariant which, in turn, led to the notoriously non-diagonal
Lagrangian matrix.  The functional
(\ref{eq:varprinconstr-dsic-static}) formulated in terms of the
double-set can now be considered again as being unitary invariant with
respect to the $\varphi_i$ because any rotation within the
$\{\varphi_i\}$ can be compensated by proper counter-rotation of the
$v_{i\alpha}$. Thus we can always perform a transformation to
the simpler functional
$$
  F[\varphi_i,v_{\alpha i}]
  =
  E_\mathrm{SIC}[\varphi_i,v_{\alpha i}]
  -
  \sum_{j}(\varphi_j|\varphi_j)\varepsilon_{j}
  -
  \sum_{\alpha\beta}
    \big(\sum_i v_{\alpha i}^*v_{\beta i}^{\mbox{}}\big)
    \Lambda_{\beta\alpha}
 \ .
$$
First, we perform variation with respect to the $\varphi_i^*$.
The key piece is
\begin{eqnarray*}
  \frac{\delta E_\mathrm{SIC}}{\delta\varphi_i^*(\mathbf{r})}
  &=&
  \sum_\alpha
  \frac{\delta\psi_\alpha^*(\mathbf{r})}{\delta\varphi_i^*(\mathbf{r})}
  \frac{\delta E_\mathrm{SIC}}{\delta\psi_\alpha^*(\mathbf{r})}
  =
  \sum_\alpha
  (\mathbf{r}|\hat{h}_\alpha|\psi_\alpha)v_{\alpha i}^*
  =
  \sum_\alpha
  (\mathbf{r}|\hat{h}_\alpha|\psi_\alpha)(\psi_\alpha|\varphi_i)
\\
  &=&
  (\mathbf{r}|\hat{h}_\mathrm{SIC}|\varphi_i)
  \ .
\end{eqnarray*}
Thus we obtain from $\delta_{\varphi_i^*}F=0$ the
first SIC equation (\ref{eq:static-diaq}).
In a second step, we perform variation with respect to the
transformation coefficients $v_{\alpha i}$. We exploit
$$
  \frac{\delta E_\mathrm{SIC}}{\delta v_{\alpha i}}
  =
  \int \textrm d^3\mathbf r \,
  \frac{\delta\psi_\alpha^*}{\delta v_{\alpha i}}
  \frac{\delta E_\mathrm{SIC}}{\delta\psi_\alpha^*}
  =
  (\varphi_i|\hat{h}_\mathrm{SIC}|\psi_\alpha)
  =
  (\varphi_i|\hat{h}_\alpha|\psi_\alpha)
$$
and obtain from variation
$
 \sum_\beta v_{\beta i}^*(\psi_\beta|\hat{h}_\alpha|\psi_\alpha)
 =
 \sum_\beta\sum_i v_{\beta i}^*\Lambda_{\beta\alpha}
$
and subsequently
$
(\psi_\beta|\hat{h}_\alpha|\psi_\alpha)=\Lambda_{\beta\alpha}
$.
Similar as in sections \ref{sec:vary-single} and \ref{sec:indepSIC},
we build the hermitian conjugate and take the difference. This yields
then the symmetry condition (\ref{eq:symcond2b}).

Thus the direct variational derivation recovered nicely the double-set
formulation of stationary SIC. It is important to remark that the
symmetry condition (\ref{eq:symcond2b}) results from minimization of
the SIC energy with orthonormality constraint for fixed $\varphi_i$.

\subsubsection{The existence of a solution to the symmetry condition}
\label{sec:symcond_sol}

The symmetry condition Eq.~(\ref{eq:symcond2b}) is as such a highly
non-linear equation. One may wonder whether a solution exists in
general. {We have seen in the above section that} the symmetry
condition simply emerges from minimizing the total energy in the
reduced space of {occupied} single-particle orbitals. There
necessarily exists an energy minimum in the restricted space and thus
that there always exists a solution to the symmetry condition.
This is a crucial feature because the symmetry condition is
always present in any formulation of SIC, static and time-dependent.

\section{Time Dependent SIC (TDSIC)}
\label{sec:TDSIC}

Now that a proper SIC formulation has been given in the static case,
we can consider the dynamical case along the same line. The diagonal
formulation will become crucial.

\subsection{Derivation of TDSIC}
\label{sec:tdsic_1set}

The TDSIC {equations are} obtained from the principle of
stationary action using the SIC energy functional (\ref{eq:fsicen})
\begin{eqnarray}
 0
 &=&
 \delta S
 \quad,\quad
 S
 =
 \int_{t_0}^{t} \textrm dt' \Big(
  E_\mathrm{SIC}
  -
 \sum_{\alpha}(\psi_\alpha|\mathrm{i}\hbar \partial_t|\psi_{\alpha})
  -
  \sum_{\beta,\gamma}^{}(\psi_{\beta}|\psi_{\gamma})
  \lambda_{\gamma\beta}
 \Big)
 \ ,
\label{eq:varprinconstr}
\end{eqnarray}
explicitly including the orthonormality constraint with Lagrange 
multipliers $\lambda_{\gamma\beta}$ as in the static case.
Note that the matrix of Lagrangian multipliers is hermitian
as shown in appendix \ref{app:hermconstr}.
Variation with respect to $\psi^*_{\alpha}$ yields the TDSIC equation
for the propagation of single-particle orbitals as
\begin{subequations}
\label{eq:tdsic}
\begin{eqnarray}
&&
  \big(
   {\hat h}_\mathrm{SIC}-\mathrm{i}\hbar\partial_t
  \big) 
  |\psi_{\alpha})
  = 
  \sum_{\beta} |\psi_{\beta})\lambda_{\beta\alpha} \quad,
\label{eq:tdsic0a}\\
&&
  \lambda_{\beta\alpha} 
  = 
  (\psi_{\beta}|h_{\alpha} - 
   \mathrm{i}\hbar\partial_t|\psi_{\alpha})
   \quad.
\label{eq:tdsic0b}
\end{eqnarray}
The relation (\ref{eq:tdsic0b}) for the Lagrangian multipliers
becomes non-trivial by the fact that it is hermitian, i.e.
\begin{equation}
  \lambda_{\beta\alpha}
  =
  \lambda_{\alpha\beta}^*
  \ .
\label{eq:tdsic0c}
\end{equation}
This can be exploited by the same steps as performed in
the static sections
\ref{sec:vary-single} and \ref{sec:indepSIC}.
We build the hermitian conjugate of Eq. (\ref{eq:tdsic0b}),
insert Eq. (\ref{eq:tdsic0c}), 
and take the difference. This yields once
again the symmetry condition
\begin{equation}
  0
  =
  (\psi_\beta|U_\beta-U_\alpha|\psi_\alpha)
  \ ,
\label{eq:symcond_td}
\end{equation}
\end{subequations}
now for TDSIC and to be fulfilled at each instant of time.

\subsection{Solution of TDSIC with a double-set of orbitals}
\label{sec:tdsic_2set}

The TDSIC equations (\ref{eq:tdsic}) are very involved and it is
extremely hard to deduce a transparent numerical stepping scheme from
them. Time evolution is related to energies and we have seen in static
SIC that single-particle energies and SIC potentials are taking
different cuts through the single-particle Hilbert space.  The concept
of single-particle energies led us naturally to a double-set strategy,
see section \ref{sec:dsetSIC}. That strategy becomes extremely
helpful in developing a solution scheme for TDSIC.

We disentangle the involved equations of motion (\ref{eq:tdsic}) by
distinguishing the SIC localizing set $\{\psi_\alpha(t)\}$ from a
propagating set $\{\varphi_i(t)\}$.  The both are connected by a
unitary transformation amongst occupied states 
}
\begin{equation}
  |\varphi_i(t))
  =
  \sum_{\beta=1}^N |\psi_{\beta}(t))\ v_{\beta i}(t)
  \quad,\quad
  \sum_\alpha v_{\alpha i}^*(t) v_{\alpha j}^*(t)
  =
  \delta_{ij}
  \;.
\label{eq:ut-td}
\end{equation}
That is the time-dependent generalization  of
the transformation (\ref{eq:unitrans}). The transformation
coefficients depend also on time and the transformation is
performed at each instant of time.

We now choose the propagating set $\varphi_i$ such that it
diagonalizes the Lagrangian matrix $\lambda_{\alpha\beta}$. Thus
we obtain 
$\left({\hat h}_{\rm SIC}-\mathrm{i}\hbar\partial_t\right)|\varphi_i)
 =
 \lambda_{ii}|\varphi_i)
$. 
The $\lambda_{ii}$ yields an irrelevant phase and can be
ignored. There remains
\begin{equation}
  \left({\hat h}_{\rm SIC} - \mathrm{i}\hbar\partial_t \right)
  |\varphi_i) 
  = 
  0
  \;,
\label{eq:diag_tdsic}
\end{equation}
The $\{ \varphi_i \}$  can then be propagated in standard
manner as:
\begin{subequations}
\label{eq:tdsic-2}
\begin{equation}
  |\varphi_i(t))
  =
  \exp{\left\{
    -\frac{\mathrm{i}}{\hbar}\int_{t_0}^{t}\, \textrm dt'\,
     \hat{h}_{\rm SIC}(t')\right\}}
  |\varphi_i(t_0)) 
  \ .
\label{eq:diag_propag}
\end{equation}
This procedure implies that the symmetry condition
(\ref{eq:symcond_td}) is fulfilled at each instant of time. To that
end, we exploit the freedom of choice of $v_{\alpha i}$. Similar as in
static SIC, we know that we can always determine the $v_{\alpha i}$
such that
\begin{equation}
  v_{\alpha i}(t)
  \quad\longleftrightarrow\quad
  0
  =
  (\psi_\beta| U_\beta-U_\alpha |\psi_\alpha)
  \ .
\label{eq:step2}
\end{equation}
\end{subequations}
The interlaced stepping of Eqs. (\ref{eq:tdsic-2})
provides a 
manageable solution scheme. Further formal properties will be
discussed in sections \ref{sec:vartdsic} and \ref{sec:csv_tdsic}.
The practical applicability and stability of the scheme is proven by
the many results presented in section \ref{sec:numgen}.

Note that the above propagator in Eq.~(\ref{eq:diag_propag}) is not
strictly unitary because $\hat{h}_\mathrm{SIC}$ is not hermitian. But
the hermiticity within occupied space, Eq.~(\ref{eq:tdsic0c}),
guarantees that the propagation (\ref{eq:diag_propag}) preserves
orthonormality within occupied space, i.e.
$(\varphi_i(t)|\varphi_j(t))=\delta_{ij}$.  That suffices for our
purposes.

\subsection{Projector notation}

An instructive alternative formulation can be given using
the operator of projection onto the unoccupied
space, the $\hat{\Pi}_\perp$ as defined in
eq. (\ref{eq:proj}).  This allows to recast the TDSIC equation into
the particularly compact form
\begin{equation}
%  \hat{\Pi}_\perp (\mathrm{i}\hbar\partial_t-\hat{h}_\alpha)
%  |\psi_\alpha) = 0 \qquad {\rm or}\qquad 
  \hat{\Pi}_\perp
  (\mathrm{i}\hbar\partial_t-\hat{h}_\mathrm{SIC})|\psi_\alpha) = 0
  \quad.
\nonumber
\end{equation}
It defines the part of the change of the wavefunctions evolving into
the space orthogonal to the already occupied states.  The evolution
inside the occupied states is again prescribed by the symmetry
condition (\ref{eq:step2}).

{
\subsection{Direct variational formulation of double-set TDSIC}
\label{sec:vartdsic}

In section \ref{sec:tdsic_1set}, we deduced TDSIC from
variation of the action  (\ref{eq:varprinconstr}) with respect to
the single-particle states $\{\psi_\alpha\}$, in a one-set strategy,
while imposing their orthonormality. The double-set TDSIC was introduced in
section \ref{sec:tdsic_2set} as a means to solve the TDSIC equations.
In this section, we are going to derive double-set TDSIC directly from
the time-dependent variational principle.  This makes the derivation
of TDSIC especially straightforward and it will add new aspects to the
scheme.

Starting point is again the action (\ref{eq:varprinconstr}), but now
formulated in terms of the two sets of orbitals, the SIC orbitals
$\{\psi_\alpha\}$ and the propagating orbitals
$\{\varphi_i\}$, related by a unitary transformation (\ref{eq:ut-td}).
Variation with imposing orthonormality of the $\{\varphi_i\}$ and
$\{v_{i\alpha}\}$ reads
\begin{eqnarray}
  S[\varphi_i,v_{i\alpha}]
  &=&
  \int_{t_0}^{t} \textrm dt' \Big(
    E_\mathrm{SIC}[\psi_\alpha]
    -
    \sum_{\alpha}(\varphi_i|\mathrm{i}\hbar \partial_t|\varphi_i)
    -
    \sum_{k,l}(\varphi_k|\varphi_j)\theta_{jk}
\nonumber\\
    &&\qquad\qquad\qquad\quad
    -
    \sum_{\alpha\beta}
    \big(\sum_i v_{\alpha i}^*v_{\beta i}^{\mbox{}}\big)
    \Lambda_{\beta\alpha}
 \Big)
 \quad.
\label{eq:varprinconstr-dsic}
\end{eqnarray}
As proven in appendix \ref{app:hermconstr}, the 
Lagrangian matrices $\theta_{jk}$ and
$\Lambda_{\beta\alpha}$ are hermitian.
Note that the transformation (\ref{eq:ut-td}) leaves the
time-derivative term invariant and we have chosen to express it in
terms of the propagating set which is here the natural choice.  The
SIC energy (\ref{eq:fsicen}) is not unitary invariant and needs to be
expressed in terms of the SIC set $\{\psi_\alpha\}$.  We ought to
remind, however, that the $\{\psi_\alpha\}$ are given through the
$\{\varphi_i\}$ via Eq.~(\ref{eq:ut-td}) such that we consider
the action as a functional of the $\{\varphi_i\}$ and 
$\{v_{\alpha i}\}$.

First, we perform variation with respect to the coefficients of the
unitary transformation. We note that, among the first three terms,
only $E_\mathrm{SIC}$ depends on them and thus $\delta_{v_{\alpha
    i}}S=0$ leads to
$$
  \delta_{v_{\alpha i}}
  \Big(
    E_\mathrm{SIC}
    -
    \sum_{\alpha\beta}
    \sum_i v_{i\alpha}^*v_{i\beta}^{\mbox{}}\Lambda_{\beta\alpha}
  \Big)
  =
  0
$$ 
and subsequently to
$$
  \int \textrm d^3 \mathbf r
  \Big(
   \underbrace{
    \frac{\partial\psi_\alpha^*}{\partial v_{\alpha i}}
   }_{
    \varphi_i^*
   }
   \underbrace{
    \frac{\partial E}{\partial\psi_\alpha^*}
   }_{
    \hat{h}_\mathrm{SIC}\psi_\alpha
   }
   -
   \sum_\beta
    v_{\beta i}^*\Lambda_{\beta\alpha}
  \Big)
  =
  0
  \quad,
$$ where $\hat{h}_\mathrm{SIC}$ is given in Eq. (\ref{eq:hsic}). That
can be rewritten in the more familiar form as
$
  (\varphi_i|\hat{h}_\alpha|\psi_\alpha)
  =
  \sum_\beta v_{\beta i}^*\Lambda_{\beta\alpha}
$
and finally be transformed to 
\begin{equation}
  (\psi_\beta|\hat{h}_\alpha|\psi_\alpha)
  =
  \Lambda_{\beta\alpha}
  \quad.
\nonumber
%\label{eq:constreq}
\end{equation}
Considering the complex conjugate of that equation and exploiting
hermiticity of $\Lambda_{\beta\alpha}$, we find
$(\psi_\beta|\hat{h}_\beta-\hat{h}_\alpha|\psi_\alpha)=0$ and from
this the crucial symmetry condition
\begin{subequations}
\begin{equation}
  v_{\alpha i}
  \quad\longleftrightarrow\quad
  (\psi_\beta|U_\beta-U_\alpha|\psi_\alpha)
  =
  0
  \quad,
\label{eq:symmcond3}
\end{equation}
which is here to be understood as a condition determining the
coefficients $v_{i\alpha}$ for given set $\{\varphi_i\}$.

In a second step, we perform variation with respect to the propagating
orbitals.  Evaluating the variational equation
$\delta_{\varphi_i^*}S=0$ yields
\begin{eqnarray}
  (\hat{h}_{\rm SIC}- \textrm i \hbar \partial_t)|\varphi_i) 
  = 
  \sum_{j}  |\varphi_j) \theta_{ji}
  \quad,\quad
  \theta_{ji} 
  = 
  (\varphi_j|\hat{h}_{\rm SIC}-\textrm i \hbar \partial_t|\varphi_i) 
%=\theta_{ij}^* 
\nonumber
\end{eqnarray}
The Lagrangian matrix is again hermitian, i.e. 
$\theta_{i j}^{\mbox{}}=\theta_{j i}^*$ 
which, in turn, implies the "weak" hermiticity condition that
$\hat{h}_{\rm SIC}$ is hermitian in the sub-space of occupied states.
Thus this Hamiltonian can be diagonalized and it is sufficient to
solve
%\begin{equation}
$
  (\hat{h}_{\rm SIC}-\textrm i \hbar \partial_t)|\varphi_i) 
  = 
  \eta_{i} |\varphi_i) 
$.
%\end{equation}
The Floquet index $\eta_{i}$ produces 
a global phase factor which is irrelevant for our purposes
and can be dropped.
Thus we remain with the time-dependent mean-field equation for
the propagating states
\begin{equation}
  (\hat{h}_{\rm SIC}-\textrm i \hbar \partial_t)|\varphi_i) 
  = 
  0
  \quad.
\label{eq:tpropag}
\end{equation}
\end{subequations}

The propagation (\ref{eq:tpropag}) together with the symmetry
condition (\ref{eq:symmcond3}) constitute the complete set of
dynamical equations for TDSIC with double set. It is satisfying to see
that both equations can be derived out of one variational
principle. The difference to TDSIC with one set as derived in section
\ref{sec:tdsic_1set} is that we now allow for two independent sets of
orbitals connected by a unitary transformation. The variational scheme
exploits that additional freedom to deliver correctly double-set TDSIC
as the optimal scheme when dealing with two sets.

We see also that the symmetry condition emerges from variation of
$E_\mathrm{SIC}$ much similar as in the static case. The reasoning
of section \ref{sec:symcond_sol} proving the existence of a solution 
for the symmetry condition does also apply here.

}

\subsection{Conservation laws}
\label{sec:csv_tdsic}

The TDSIC equations yield energy conservation as long as the external
field remain independent on time. And they also fulfill the zero-force
theorem \cite{Mun07a,Lev85a,Vig95a}. The reasoning is simple. The LDA
functional and its SIC extension  are invariant under time- and
space-translations. The same holds for the orthonormality
constraints. The equations of motion are derived variationally without
further restrictions and approximations. This yields energy
conservation from time translational invariance. 
The space-translational invariance would yield momentum-conservation
if the electrons were alone. That feature is broken by external
fields. But what remains is the fact that the electrons cannot exert a
force on themselves which is the content of the zero-force theorem.
More explicit proofs are given in appendix \ref{sec:conserv_laws}.

\subsection{Static limit}

It is also interesting to consider the stationary limit of TDSIC. 
To that end, we use TDSIC in double-set set formulation. 
We identify
$$
  \varphi_i(\mathbf{r},t)
  =
  \varphi_i(\mathbf{r,0})e^{- \textrm i \epsilon_i t/\hbar}
  \ .
$$ 
Inserting that into 
the TDSIC equation (\ref{eq:diag_tdsic}), one immediately 
recovers the static SIC equation (\ref{eq:static-diaq})
for the diagonalizing states.
The single-particle wavefunctions change in time only by a phase
factor. The sub-space of occupied states thus remains constant in
time and the symmetry condition always minimizes the same sub-space.
And that yields the time-independent localized states $\psi_\alpha$
of the static problem.

\section{Test cases and numerical realization}
\label{sec:numgen}

\subsection{Models}
\label{sec:models}

We want to test the above discussed TDSIC formalism in truly dynamical
situations. The ultimate goal of these studies is to describe
dynamically the irradiation of various molecules, in particular
organic ones.  {A correct modeling of the IP} is then crucial
and TDSIC {becomes} compulsory, especially close to emission
threshold. We have developed since long fully fledged coupled
electronic dynamics for clusters \cite{Cal00} and used extensively at
TDLDA level and in simplified SIC schemes \cite{Leg02}. We thus have
at hand a powerful tool for analyzing irradiation dynamics. Still, the
proposed TDSIC formalism is quite involved. In order to have a more
flexible testing tool, we have thus developed a one-dimensional (1D)
model mocking up typical atoms and simple (linear !) molecules. This
will serve as a starter to study the properties of TDSIC. And we will
finally complement these schematic results by realistic ones with the
full 3D approach.

\subsubsection{The simplified 1D model}
\label{sec:1D}

To test numerically the SIC scheme, we take up
the test case of \cite{Hen98} consisting in a one-dimensional
model for a molecule. Spin is not taken into account explicitly
and all electrons are assumed to have the same spin such that they
all explore the full exchange effects.
Apart from its simplicity and computational
cost, the 1D test case has also the advantage to be a much more sensitive
{test of} orthonormalization than in 3D calculations.

For the electron-electron interaction, we use a smoothed Coulomb potential (in Hartree units)
\begin{equation}
v(x,x')=\frac{1}{\sqrt{(x-x')^2+a}} \quad.
\end{equation}
Starting with this "elementary" interaction, we construct the
corresponding LDA energy functional for exchange only. Working
at the level of exchange only allows to have fully fledged
time-dependent HF (TDHF) calculations as a benchmark to which TDSIC
calculations can be compared.  The detailed calculation of the LDA
energy is presented in appendix \ref{sec:lda1d}. The resulting LDA
exchange potential ($\gamma$ is the possible degeneracy number, equal
to 1 in the next results) reads:
\begin{equation}
  U_{\rm LDA}^x[\rho]
  =
  -\frac{1}{2\pi}\int_{-\infty}^{+\infty} \textrm dy 
   \frac{\sin(\frac{2\pi\rho(x)}{\gamma}y)}{y\sqrt{y^2+a}} \quad.
%\nonumber\\
%&&U_{LDA}^x[\rho_i](x)
%=
%-\frac{1}{2\pi}\int_{-\infty}^{+\infty} dy 
%\frac{\sin(\frac{2\pi\rho_i}{\gamma}y)}{y\sqrt{y^2+a}}
\label{eq:1Dpot}
\end{equation}

For the electron-ion interaction, we also use a smoothed Coulomb
potential (in Hartree units) of the form 
\begin{equation}
U_{\rm ion}(x)=-\frac{N z}{\sqrt{(x-R/2)^2+b}}-\frac{N 
  (1-z)}{\sqrt{(x+R/2)^2+b}} \quad,
\end{equation}
where $R$ is the inter-ionic distance which allows the possibility to
compute bi-atomic molecules, $N$ is the number of electrons and $z \in
[0,1]$ is an asymmetry parameter. If $z=0.5$ and $R$=0, we recover
the atomic case. If $z$=0.5 and $R\ne$ 0, we obtain the bi-atomic
symmetric molecular case. If $z\ne$ 0.5 and $R\ne$ 0, we get the
asymmetric molecular case, which provides a more critical
probe of the role of orthonormalization.  

The $a$ and $b$ parameter for a two-electron system are scaled to
approach the experimental H$_2$ bond length (1.4 $a_0$) and ionization
potential (0.57 Ha). With $a$=0.8 and $b$=0.5, we obtain a H$_2$ bond
length of 1.6 a$_0$ and an ionization potential of 0.5 Ha.  
Depending on the case, we use slightly varied values
which will be indicated.

\subsubsection{Full 3D model}
\label{sec:3d}

The 3D calculations follow the standard techniques as we use it since
long \cite{Rei03a,Cal00}. The electron wavefunctions and spatial
fields are represented on a Cartesian grid in three-dimensional
coordinate space.  The spatial derivatives are evaluated via fast
Fourier transformation. The ground state configurations were found by
adapting the accelerated gradient iteration for the electronic
wavefunctions \cite{Blu92} to SIC. Propagation is done by the
time-splitting method for the electronic wavefunctions~\cite{Fei82}
augmented by updates of the symmetry condition as explained below.
For the energy functional, we employ the widely used functional
of \cite{Per92}.

\subsection{Solution scheme for stationary SIC}
\label{sec:propag_num}

In straightforward generalization of the damped gradient step
\cite{Blu92}, we solve the static Eqs. (\ref{eq:SICeqs})
iteratively as
\begin{subequations}
\begin{eqnarray}
  \psi_\alpha^{\rm(new)}
  &=&
  {\cal O}\left\{
   \psi_\alpha
   -
   \frac{\delta_{\rm step}}{\hat{T}+E_{\rm damp}}
   \left[
      \hat{h}_\alpha\psi_\alpha
      -
      \sum_\beta \psi_\beta\lambda_{\beta\alpha}
   \right]
  \right\}
  \quad,
\label{eq:dampgrad}
\\
  &&
  \lambda_{\beta\alpha}
  =
  \frac{(\psi_\beta|\hat{h}_\alpha|\psi_\alpha)
        \!+\!
        (\psi_\alpha|\hat{h}_\beta|\psi_\beta)^*}
       {2}
  \quad,
\label{eq:lamda-comp}
\end{eqnarray}
\end{subequations}
where ${\cal O}$ stands for orthonormalization. The step
(\ref{eq:dampgrad}) provides simultaneously a complete solution
including a matching of the symmetry condition, because the latter
is implicitly taken into account in the symmetrization of the
$\hat\lambda$ matrix.

It turns out that the step (\ref{eq:dampgrad}) converges,
however 
very slowly, and that the symmetry condition causes the delay.  We
speed up the iterations improving the symmetry condition explicitly
in each step. This is done by a unitary transformation within occupied
states. The coefficients of that unitary transformation are also
determined by  a gradient iteration as
\begin{subequations}
\begin{eqnarray}
  u_{i\gamma}^{\rm (new)} 
  &=&
  {\cal O}\left\{ u_{i\gamma}^{\rm (old)} - \eta D_{i\gamma}\right\}
  \quad,
\label{eq:symconditerUT}
\\
D_{i\gamma} &=& \partial_{u_{i\gamma}^*} 
  \left[
    E_\mathrm{SIC}
     -
    \sum_{j,\alpha,\beta}u_{j\alpha}^*u_{j\beta}^{\mbox{}}\lambda_{\beta\alpha}
  \right] 
  = 
  - (\varphi_i|U_\gamma|\psi_\gamma)
  - \sum_\beta u_{i\beta}^{\mbox{}}\lambda_{\beta\gamma}
  \quad,
\label{eq:symconditerUT2}
\end{eqnarray}
\end{subequations}
where the ``driving force'' $D_{i\gamma}$ is obtained by variation of
the SIC energy with respect to $u_{i\gamma}^*$.
That interlaced combination of damped gradient step and symmetry
condition converges acceptably fast.

Depending on the initial conditions, it may take a while until the
interlaced iteration has found its path to the properly localized
wavefunctions. A further substantial acceleration can be achieved by
performing in the initial phase once in a while a localization
transformation. There are several localization criteria at
choice \cite{Fos60,Edm63}.
We found very efficient improvements with simply minimizing the sum of
the spatial variances of the single-electron states, defined as
\begin{equation}
\Delta \psi 
= \sum_\alpha\left[
(\psi_\alpha|\mathbf{r}^2|\psi_\alpha)
-
(\psi_\alpha|\mathbf{r}|\psi_\alpha)^2
\right].
\label{eq:variance}
\end{equation}

\subsection{Propagating TDSIC numerically}
\label{sec:propag_tdsic}

In the time-dependent case, the only {manageable}  way to
propagate TDSIC is to use the double-set strategy.
For one time step $\delta t$, the propagation proceeds as
follows. We first evaluate 
$
\displaystyle
\varphi_i \left(t+\frac{\delta t}{2} \right) = \exp \left[ - \frac{
    \textrm i\, \delta t}{2 \hbar} \, \hat{h}_{\rm SIC}(t)\right]
\varphi_i(t)
$, 
where $\hat{h}_{\rm SIC}(t)$ is obtained by the chain~:
\begin{equation}
 \varphi_i(t)
 \xrightarrow{\textrm{Eqs.(\ref{eq:symcond_td},\ref{eq:ut-td})}}
 \{\upsilon_{i\beta}(t), \psi_\alpha(t)\} 
 \xrightarrow{\textrm{Eq.(\ref{eq:hsic})}} \hat{h}_{\rm SIC}(t).
\label{eq:num}
\end{equation}
Gradient iteration, similar to Eq.~(\ref{eq:symconditerUT2}), is used
to solve Eq. (\ref{eq:symcond_td}) for the $\upsilon_{i\beta}(t)$,
from which one deduces the $\psi_\alpha(t)$ and $\hat{h}_{\rm
  SIC}(t)$. The $\varphi_i(t+\delta t/2)$ thus obtained are used
to compose $\hat{h}_{\rm SIC}(t+\delta t/2)$ similarly using the chain
(\ref{eq:num}).
We finally compute 
\begin{equation}
\varphi_i(t+\delta t) = \exp \left[ - \frac{\textrm i\,
\delta t }{\hbar} \, \hat{h}_{\rm SIC} \left( t+ \frac{\delta t}{2}
  \right) \right] \varphi_i(t).
\label{eq:csicprop-perturb}
\end{equation}

After all, the scheme as explained here works reliably and
robust in 1D as well as in 3D.  We checked conservation of energy and
orthonormality and found it fully satisfying in all cases. 
The symmetry condition is, of course, fulfilled all along the time
evolution by construction.

A proper computation of ionization requires to prevent
reflection of electrons which have been ejected from the molecule and
are now impinging on the bounds of the numerical grid. We do that by
employing boundary conditions.  To that end, an absorbing zone of a
few grid points is defined.  For each time step, the part of
electronic wavefunctions which has penetrated into that zone is
eliminated by a mask function, for details see \cite{Rei03a,Cal00}.

\section{Results and discussion}

\subsection{Stationary state}
\label{sec:local}

\begin{figure}[htbp]
\begin{center}
\includegraphics[width=6.5cm,height=5.5cm,angle=0]{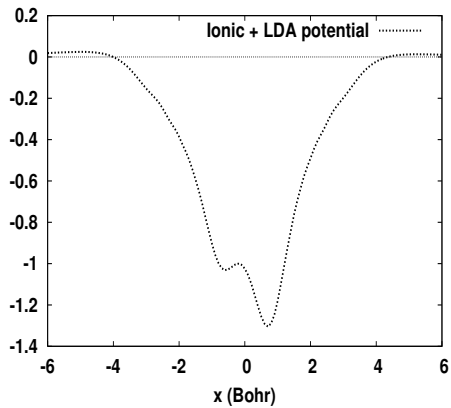}
\caption{
 The mean-field potential from ionic background and LDA-part
  (without SIC-parts) obtained in the 1D model with 2 electrons.
  The model parameters are 
  $a=0.8\,\mathrm{a}_0$, $b=0.5\,\mathrm{a}_0$, $R=1.5\,\mathrm{a}_0$, 
  and $z=0.4$.
\label{fig:1D2e.3pot}}
\end{center}
\end{figure}
It is well known that static SIC has a tendency to localize
spatially the orbitals \cite{Ped84}. We analyze this fact on the
example of the stationary solution for a system of two electrons in
the 1D model developed in section \ref{sec:1D}.  
Note that  $z\neq 0.5$, i.e. we enforce a slight asymmetry.
The resulting ionic + LDA potential obtained for a SIC solution is
plotted in Fig. \ref{fig:1D2e.3pot}. Note that the SIC potential
cannot be plotted easily because of its state-dependence.
The resulting single-particle energies $\epsilon_i$ are
$-0.88$ / $-0.32$ for LDA, $-1.18$ / $-0.60$ for SIC, and $-1.24$ /
$-0.55$ for the HF benchmark. Note that the results of SIC comes
close to HF as it should be.

\begin{figure}[htbp]
\begin{center}
\includegraphics[width=6.5cm,height=5.5cm,angle=0]{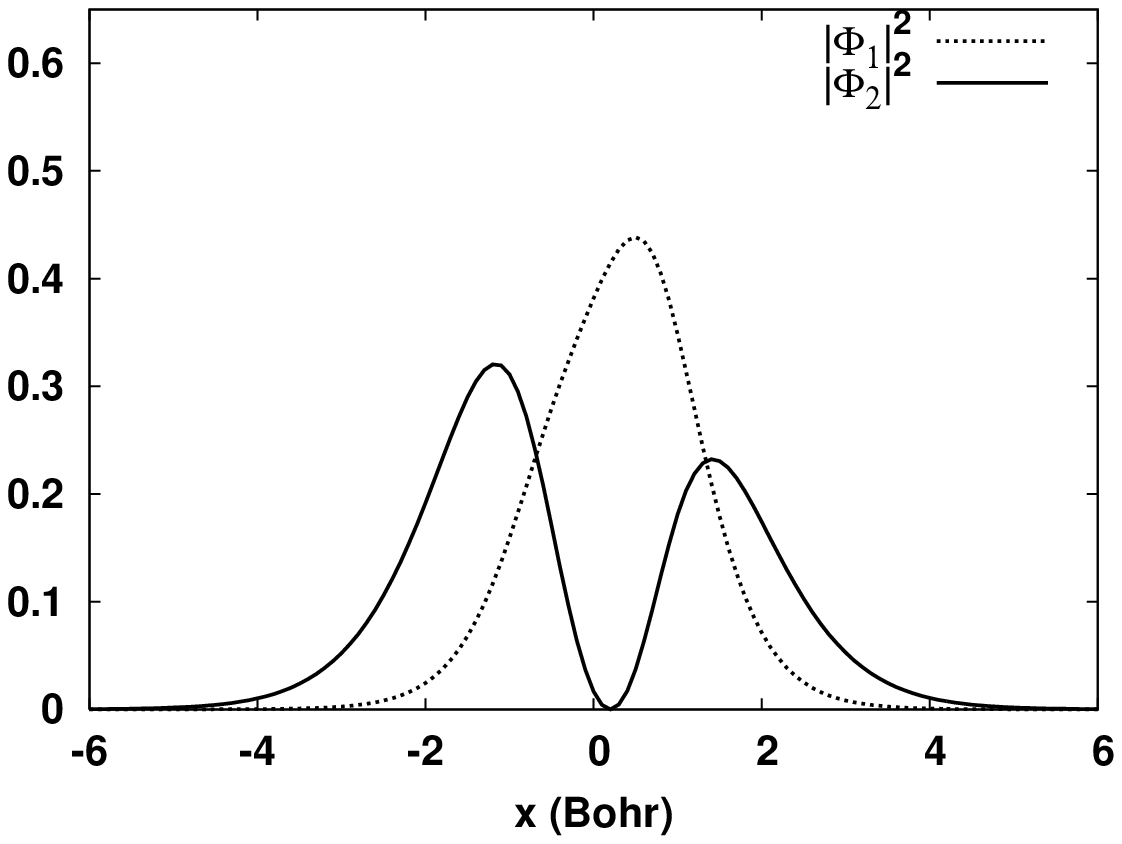}
\includegraphics[width=6.5cm,height=5.5cm,angle=0]{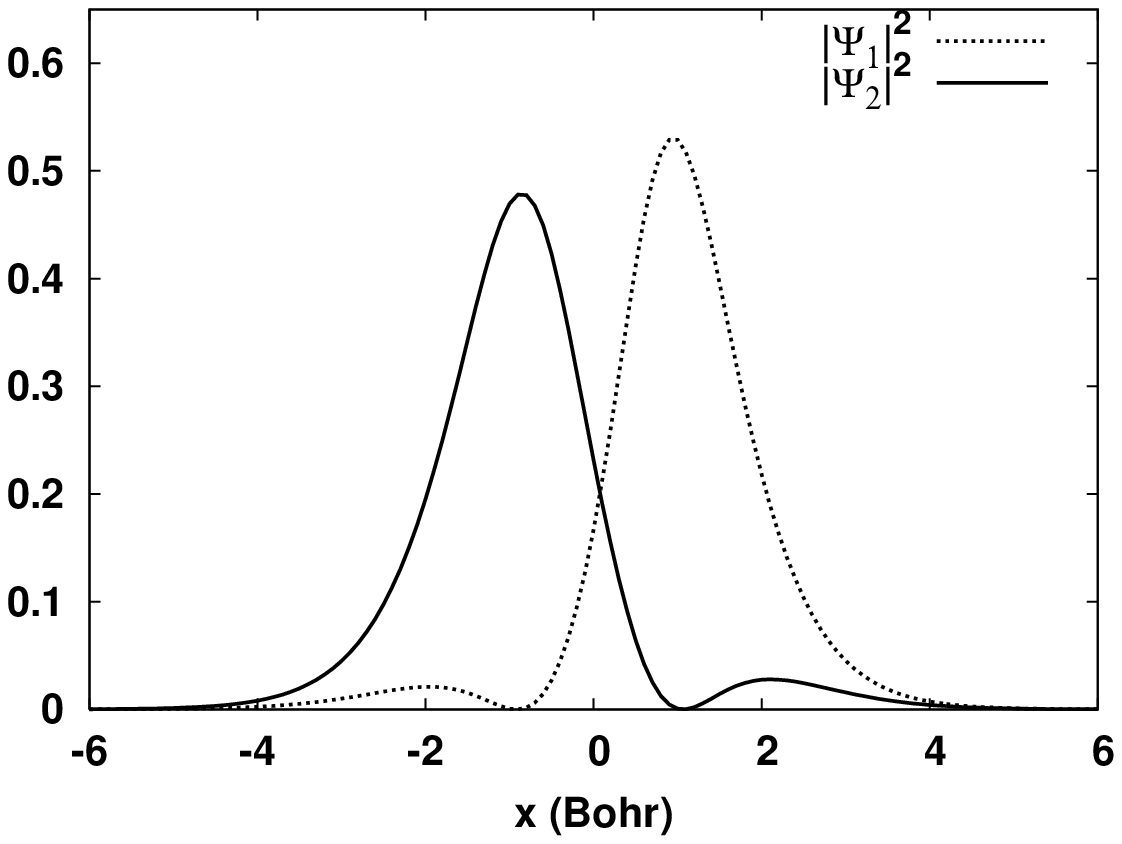}
\includegraphics[width=6.5cm,height=5.5cm,angle=0]{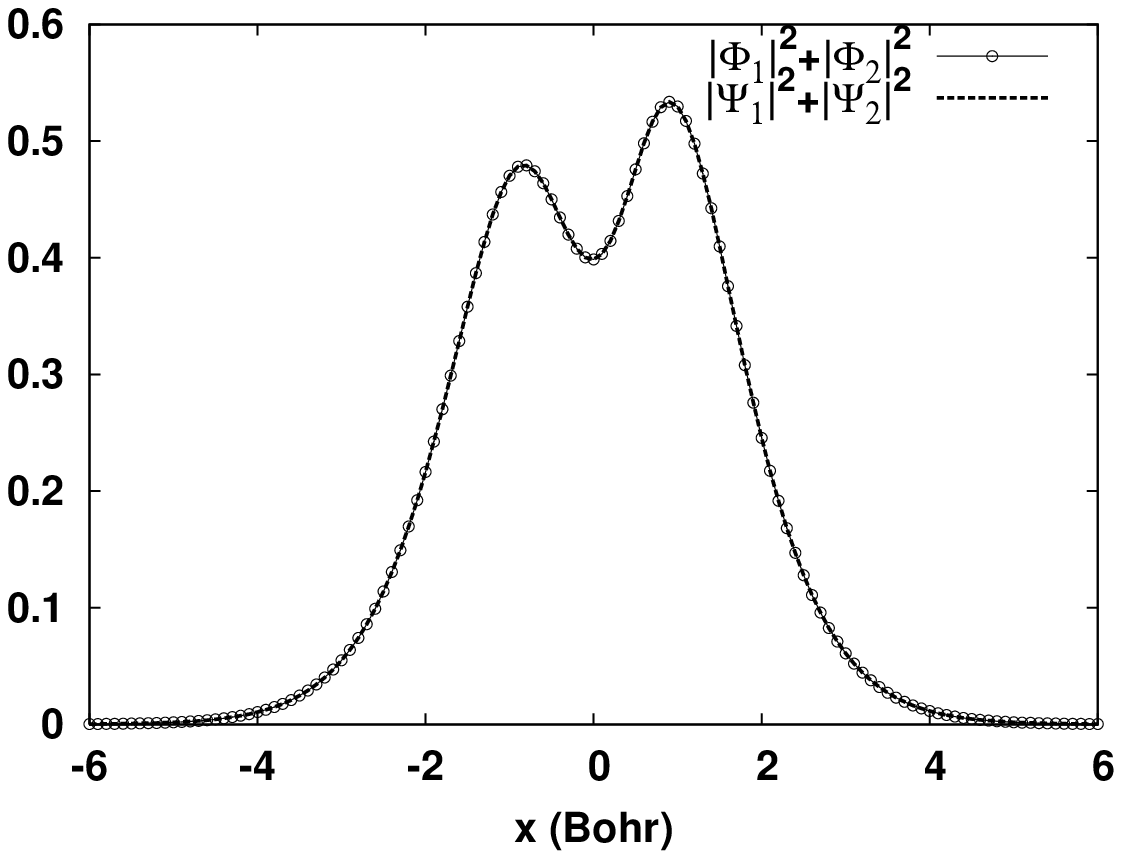}
\caption{Static SIC densities obtained in the 1D model with 2
    electrons. 
    Top left: Densities from diagonal wavefunctions, $|\varphi_i|^2$.
    Top right: Densities
    from localized wavefunctions, $|\psi_\alpha|^2$, obtained
    after unitary transformation. 
    Bottom: Comparison of the total
    density calculated from both sets of wavefunctions.
\label{fig:1D2e.3stat}}
\end{center}
\end{figure}
Fig. \ref{fig:1D2e.3stat} compares the single-particle densities of
the localized wavefunctions $|\psi_\alpha|^2$ with those of the
energy-diagonal wavefunctions $|\varphi_i|^2$. It is obvious 
that the localized densities
are much better
concentrated in space than the diagonal ones.
%  It is the minimization of SIC energy which induces
%the trend to localized single particle states $|\psi_\alpha)$.
%
The total density (lower panel), of course, remains the same for both
cases because the two sets are linked by a unitary transformation.

\begin{figure}[htbp]
\begin{center}
\includegraphics[width=12cm,height=10cm,angle=0]{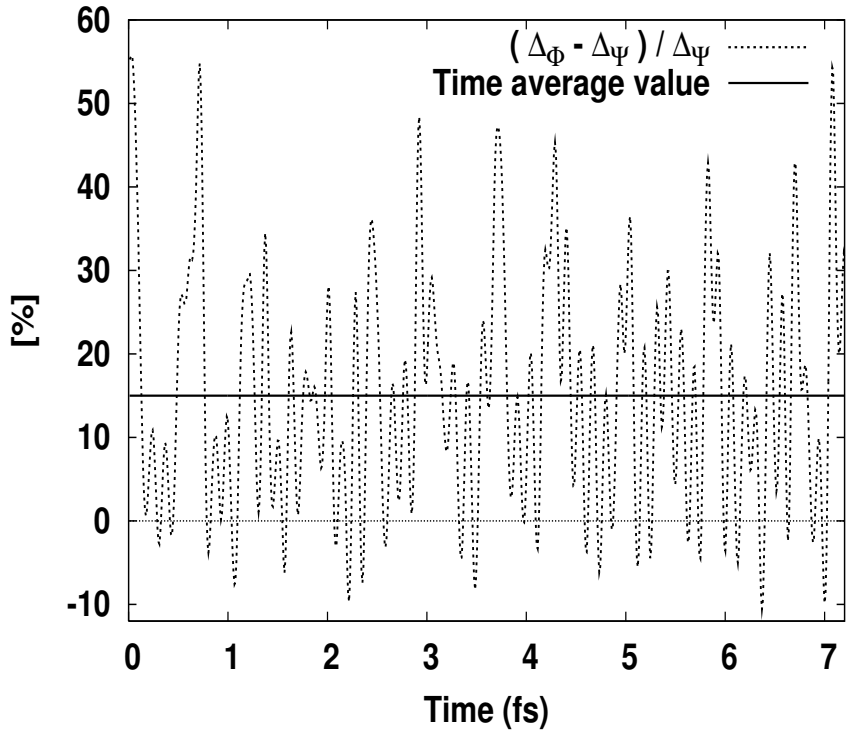}
\caption{\label{fig:1D2e.3td} 
Time evolution of single-particle variances (\ref{eq:spvariance})
obtained in the 1D model with 2 electrons and
asymmetrical ionic background. 
Shown is the 
relative value $(\Delta\varphi-\Delta\psi)/\Delta\psi)$.
The time averaged result is indicated by a straight line.
}
\end{center}
\end{figure}
%
%% \footnote{\JMcomm{In Fig. \ref{fig:1D2e.3td}, it is the $\Delta\psi$... and not the $\Delta_\psi$... which are plotted, according to
%% the notations of equations (\ref{eq:variance}) and (\ref{eq:spvariance}) - the $\sqrt()$ is not take into account. There is a mistake here.}}

We are now considering dynamical evolution from the given
static state, where the excitation is initialized by a very short
laser pulse, simulated as an instantaneous boost~\cite{Cal97b}.  The
question is to what extent the localization may be washed out through
the excitation.  In order to follow an evolution, one needs to
characterize localization by one number and we do that by the spatial
variance of a single particle states
\begin{equation}
  \Delta\varphi
  =
  \sum_i(\varphi_i|x^2|\varphi_i)-
        \big(\sum_i(\varphi_i|x^2|\varphi_i)\big)^2
\label{eq:spvariance}
\end{equation}
and similarly for $\Delta\psi$, see Eq.~(\ref{eq:variance}).  Fig.
\ref{fig:1D2e.3td} shows
the relative variance for the
propagating state $\varphi$ as compared to the localized state $\psi$
(which was starting from the diagonal stationary state).
The relative variance undergoes large fluctuations, as the
variances as such do as well. But in the average, we see the
expected result. 
The $\Delta\varphi$ remains larger than $\Delta\psi$, by about 15~\%
on time average.

\subsection{Ionization properties}
\label{sec:exs}

\begin{figure}[htbp]
\begin{center}
\includegraphics[width=6.5cm,height=5.5cm,angle=0]{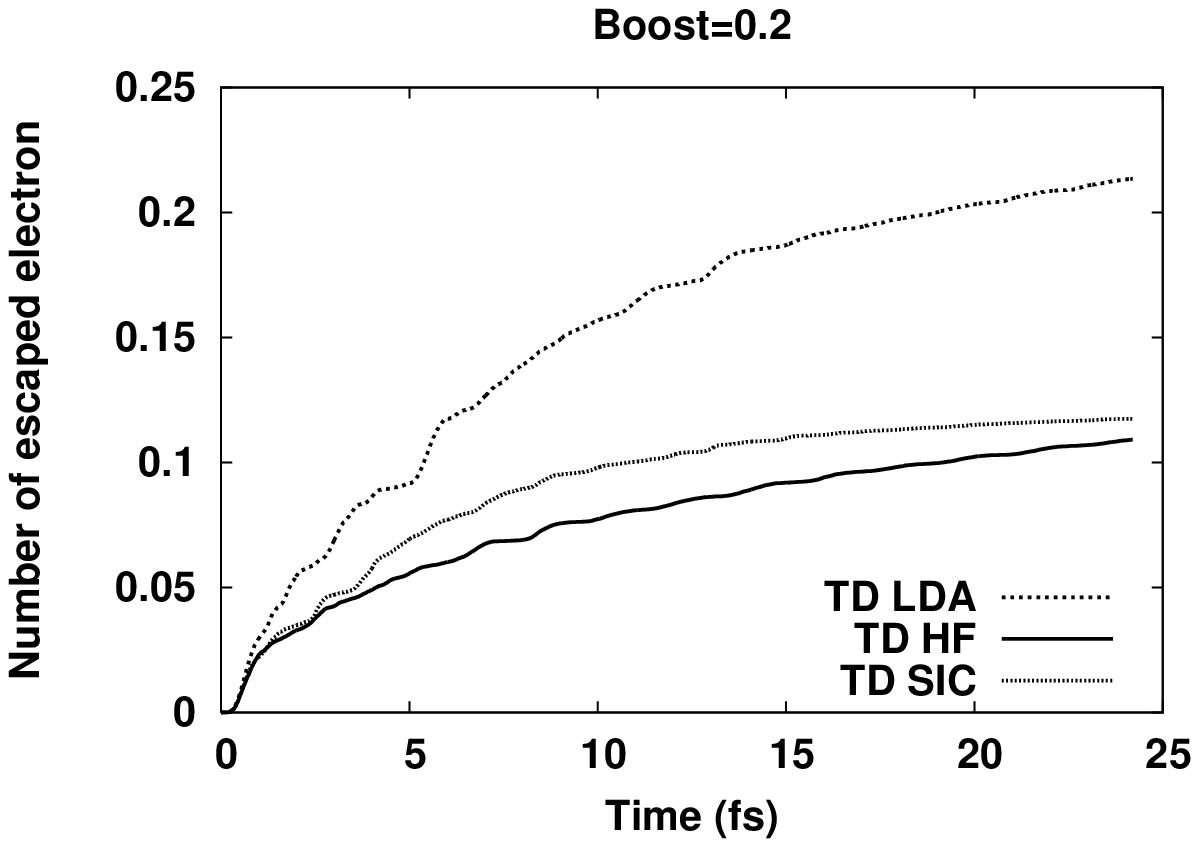}
\includegraphics[width=6.5cm,height=5.5cm,angle=0]{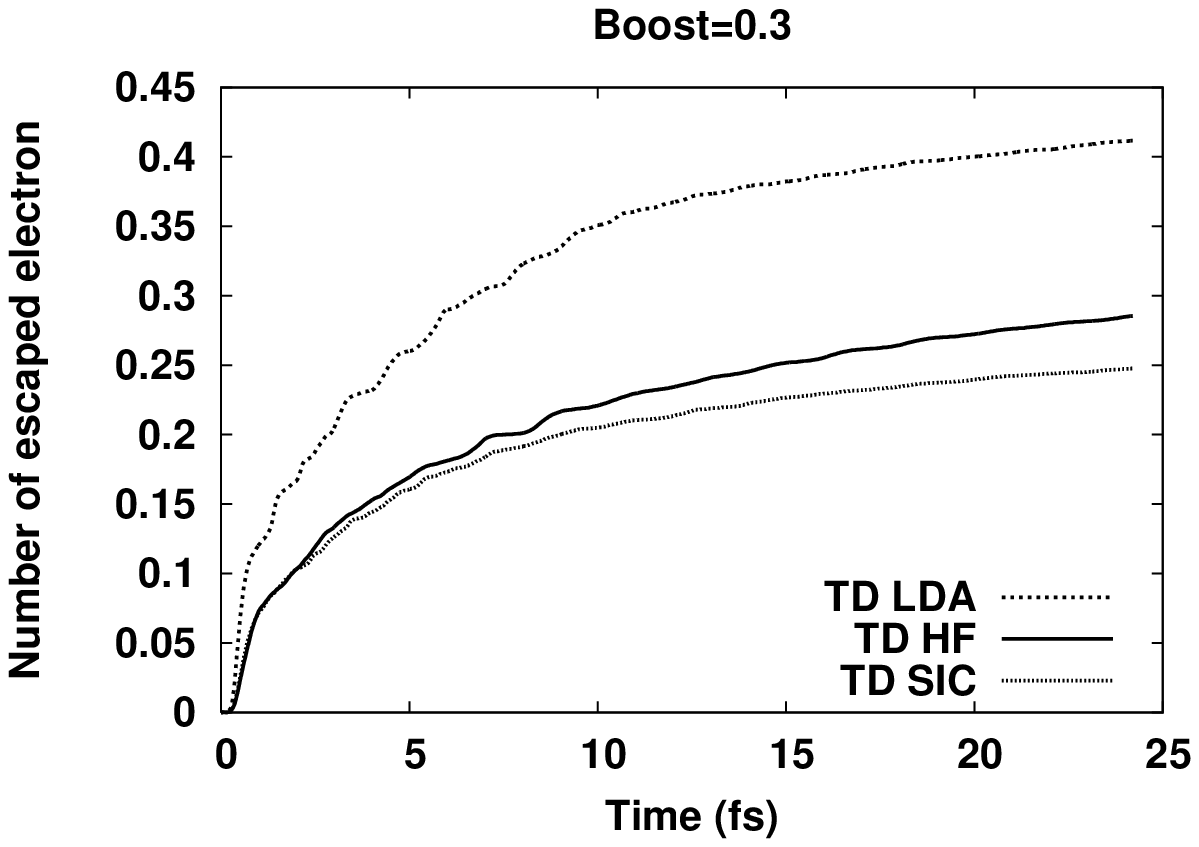}
\caption{
Number of escaped electrons, as function of time, for an 1D atom
of 3 electrons, initially excited by two different
values of boost, as indicated. Three time-dependent schemes are
compared~: TDLDA, TDHF and TDSIC.
The model parameters are $a=0.5\,\mathrm{a}_0$,
$b=0.5\,\mathrm{a}_0$, $R=0$, and $z=0.5$.
\label{fig:3e1D}
}
\end{center}
\end{figure}
As a further observable, we consider the degree of ionization which,
as stated above, is a sensitive quantity to probe the effect of SIC.
Remind that absorbing bounds are applied for a proper
handling of ionization.  We use again an instantaneous boost to
simulate a very short laser pulse. Fig.~\ref{fig:3e1D} shows the
results for an atom of 3 electrons, within the 1D model
and for two different initial boosts, comparing the TDHF benchmark
with TDLDA and TDSIC. 
The deposited energy with a boost of 0.2 / 0.3 Ha represents 60 $\%$ / 134 $\%$ of the LDA ionization energy (0.100 Ha)
and 21 $\%$ / 47 $\%$ of the SIC ionization energy (0.133 Ha).
Therefore we expect a strong ionization overestimation for LDA.
It is obvious that TDSIC comes much closer to
the benchmark (TDHF) than TDLDA.  We checked various other 1D
molecular systems and found similar results confirming that TDSIC
recovers nicely the proper ionization features.

\subsection{Results from 3D calculations}
\label{sec:3D}

\begin{figure}[htbp]
\begin{center}
\includegraphics[height=5cm]{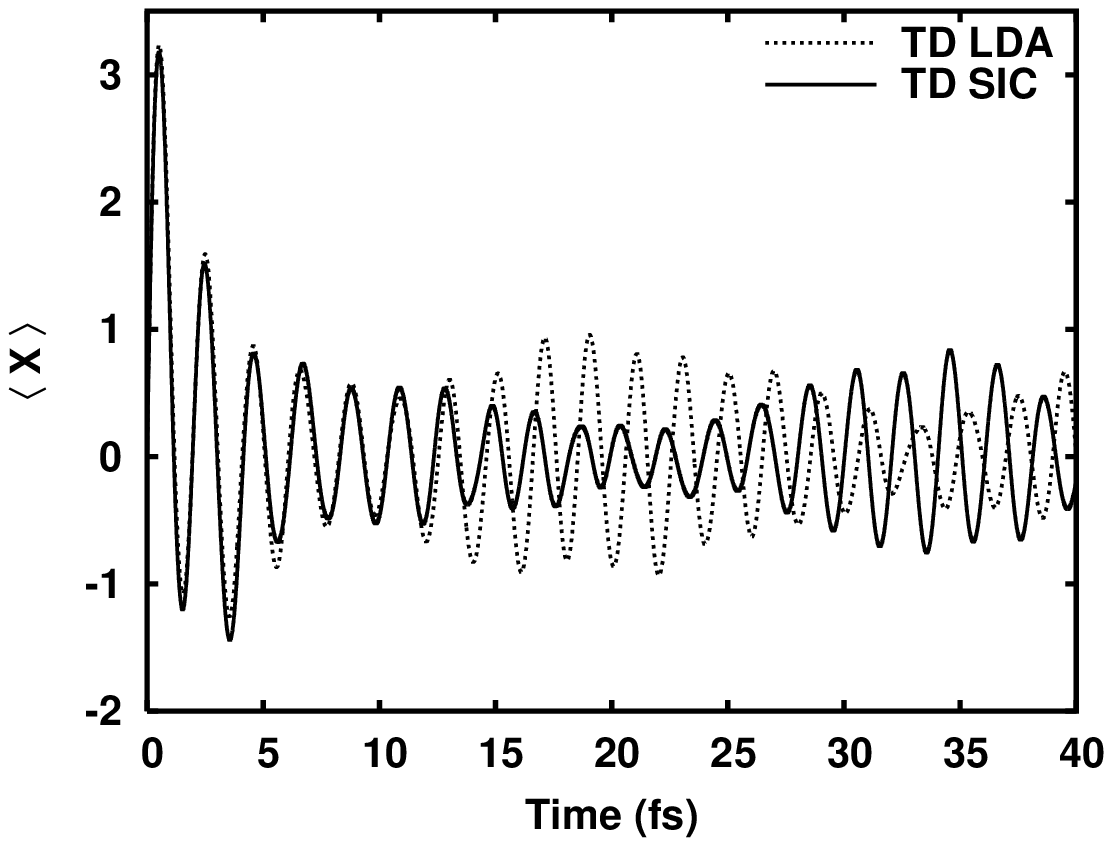}
\quad
\includegraphics[height=5cm]{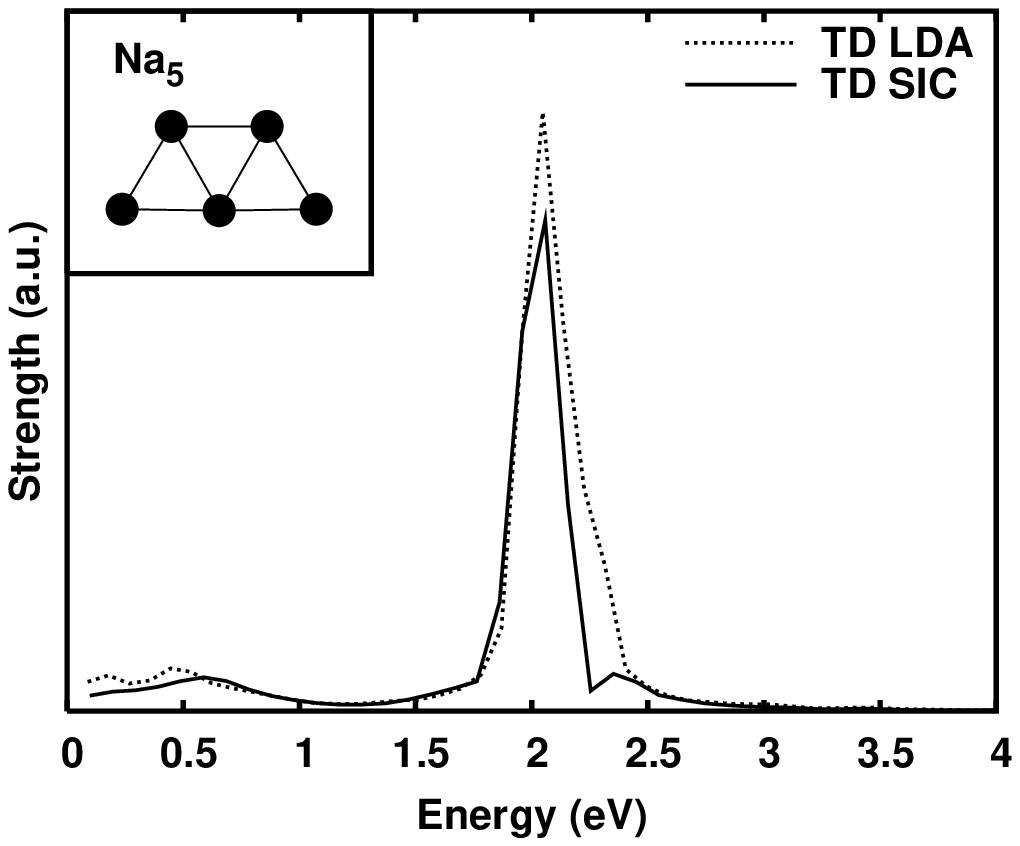}
\caption{
Na$_5$ initially excited by an instantaneous boost and computed in
full 3D, where TDSIC and TDLDA are compared.
Left~: Time evolution of the dipole spectra in $x$ direction. 
Right~: Optical absorption spectrum (in arbitrary units) derived
thereof \cite{Cal97b}; the insert shows the geometry of Na$_5$.}
% nabsorb=4.
\label{fig:Na5dipX}
\end{center}
\end{figure}

Finally, we want to check the effect of SIC in a realistic
3D situation. To that end, we consider the Na$_5$ cluster 
which has a non-symmetric planar structure (see insert in
  Fig.~\ref{fig:Na5dipX}) and whose electron cloud is weakly bound
and so provides a critical test case for formal developments
\cite{Mun07a}. This cluster contains altogether five valence 
electrons which are active in the low frequency irradiation
processes. The core electrons of the Na atoms are much more bound and
are eliminated by using pseudo-potentials.  For the electronic exchange
and correlations, we employ the energy-density functional of
\cite{Per92}.  The laser excitation is again simulated by an
instantaneous boost.  For the further details of the 3D calculations
see, e.g., \cite{Rei03a,Cal00}.
The first principle result to be mentioned is that the newly developed
solution scheme for TDSIC runs smoothly also for the full 3D case. 
%A TDHF calculations for comparison does not exist.

\begin{figure}[htbp]
\begin{center}
\includegraphics[width=9cm,height=6cm,angle=0]{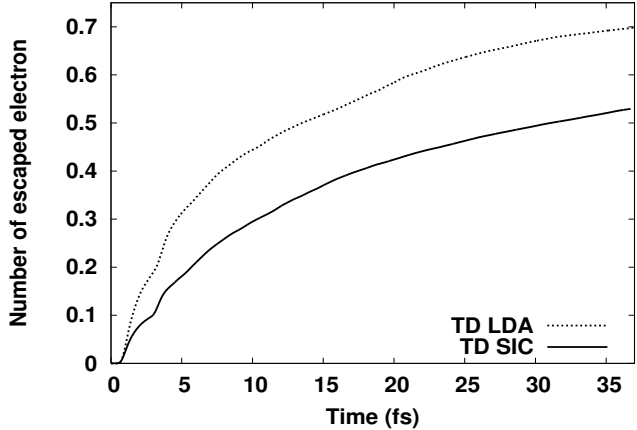}
\caption{\label{fig:Na5}
Time evolution of the number electrons escaped from the
Na$_5$ cluster,
computed in full 3D, The system was initially excited by an
instantaneous boost. TDSIC and TDLDA are compared.
% nabsorb=4.
}
\end{center}
\end{figure}
Fig. \ref{fig:Na5} compares the time evolution of ionization
between TDLDA and TDSIC, for an initial boost of $0.125$ Ha, 
whose deposited energy represents 149 $\%$ of the LDA ionization energy ($0.105$ Ha)
and 90 $\%$ of the SIC ionization energy ($0.172$ Ha). 
We see a similarly dramatic effect from SIC
on the ionization. TDSIC produces less in accordance with the fact
that SIC enhances the IP from the LDA value of $0.105$ Ha to the SIC
value  $0.172$ Ha. 

Fig. \ref{fig:Na5dipX} compares the time evolution of the
dipole oscillation as such (left) and the subsequent optical
response (right).  The time evolution (left panel) shows that the
initial oscillations are very much the same for TDLDA and TDSIC.  Some
deviations develop in the further course of propagation which emerges
from different interferences with particle-hole states.  This lets us
expect that the Mie plasmon position is not affected by TDSIC, but
that detailed fragmentation pattern may be different.  These two
features are indeed nicely found in the optical absorption spectra
(right panel).
Both results, reduced ionization together with little influence on the
dominant Mie plasmon excitation was also found in earlier studies
using a simplified SIC scheme, time-dependent average-density SIC
\cite{Leg02,And02b}.

%% \begin{figure}[htbp]
%% \begin{center}
%% \includegraphics[width=9cm,height=6cm,angle=0]{Cb.6esc.eps}
%% \caption{\label{fig:C}
%% Time evolution of  the number electrons  escaped from the C atom
%% computed in full 3D, The system was initially excited by an
%% instantaneous boost. TDSIC and TDLDA are compared.
%% }
%% \end{center}
%% \end{figure}
%
While Na$_5$ is a soft system, as an
alternative example of a strongly bound system, we consider the case
of a C atom.  The IP is now $0.224$ Ha for LDA, whereas it is enhanced
to $0.452$ Ha for SIC. 
%% Fig. \ref{fig:C} compares the time evolution of ionization
%% between TDLDA and TDSIC for the C atom.  We again see a strong
%% effect from SIC, which produces much less ionization because
%% of its much enhanced IP.
Then TDSIC produces much less ionization than TDLDA when the C atom is
excited by an instantaneous boost, similarly to the case of Na$_5$
(see Fig.~\ref{fig:Na5}).
We also compute the optical response spectrum for C, although this
observable is less relevant for this non-metallic example than for
Na$_5$. We observe (not shown here) a shift between the SIC peak and
the LDA peak. SIC also makes an effect on optical absorption because
the deeper binding restricts the dipole oscillations more tightly than
in case of LDA. 

%% \begin{figure}[htbp]
%% \begin{center}
%% \includegraphics[width=6.5cm,height=5.5cm,angle=0]{Cb.6dipX.eps}
%% \includegraphics[width=6.5cm,height=5.5cm,angle=0]{Cb.6OptResp.eps}
%% \caption{
%% Left~: 
%% Time evolution of the $x$-dipole spectra for the C atom
%% initially excited by an instantaneous boost and
%% computed in full 3D .
%% Right~: The optical absorption spectrum derived thereof
%% \cite{Cal97b}.
%% TDSIC and TDLDA are compared.
%% \label{fig:C_OptResp}}
%% \end{center}
%% \end{figure}

%% Fig. \ref{fig:C_OptResp} compares time evolution
%% and optical response spectra
%% %obtained by a Fourier transform into the frequency domain.
%% for the case of the C atom. Here, we
%% observe a shift between the SIC peak and the LDA peak.  SIC
%% makes an effect also on optical absorption because the deeper binding
%% restricts the dipole oscillations more tightly than in case of LDA.

\section{Conclusion}

In that paper, we have investigated the time-dependent
self-interaction correction method (TDSIC) which augments the
time-dependent local-density approximation (TDLDA) by a
self-interaction correction (SIC). That correction becomes crucial
when aiming at the description of highly dynamical processes where
electron emission plays a role. But SIC raises problems because the
resulting one-body Hamiltonian becomes explicitly orbital dependent
which, in turn, can destroy the necessary orthonormality of the
single-particle states. One needs to add an explicit constraint on
orthonormality. This leads to a ``symmetry condition'', in other
words, to the (plausible) condition that the SIC mean-field
Hamiltonian is hermitian within the space of occupied states.  An
implementation of that involved condition in TDSIC has been achieved
by dealing with two different sets of single-particle orbitals~:
Propagating orbitals which are carried forth by standard mean-field
stepping methods and localizing (or SIC) orbitals which are used to
compute the SIC mean-field. The relation between the two sets is
established by a unitary transformation which is determined such that
the localizing set satisfies the crucial symmetry condition at each
time. The newly developed representation of TDSIC constitutes a
formally consistent and numerically reliable scheme. Crucial
conservation laws (energy, orthonormality, zero-force theorem)
are all obeyed formally and in practical calculations.

First tests have been performed in a one-dimensional model for a
molecule.  By the standard rules of LDA, we have developed a LDA
functional for exchange only. That allowed direct comparison with
(time-dependent) Hartree-Fock (TDHF) as a benchmark.  The dynamical
evolution was initiated by exciting the ground state with a very short
laser pulse. The pulse was idealized as an instantaneous boost and we
computed for various excitation strengths to check the robustness
of the results. It was found that TDSIC compares very well with TDHF,
while TDLDA overestimates ionization by 50--100\%.
Tests had also been done for fully three dimensional calculations
considering a Na$_5$ molecule as test cases.  We found again a
substantial overestimation of ionization for TDLDA, in that case by
about 30\%.

Although full 3D calculations have proven to be feasible and stable,
the TDSIC scheme is numerically costly.  We consider is as a starting
point for further developments towards more efficient schemes. 
%The SIC
%hamiltonian form (\ref{eq:hsic}) clearly shows it's non locality.  But
%as the $\psi_\alpha$ orbitals should remain localized, this
%hamiltonian is only weakly non-local. 
The most costly detail is the fulfillment of the
symmetry condition which basically provides more localized
single-particle states. 
This suggests to replace the symmetry condition
by a direct localization condition
which will be numerically less costly. A promising option to find it
is the time-dependent optimized effective potentials formalism
\cite{Kue07,Ull95a}. It has already been used in the static case,
leading to what we called a ``Generalized Slater''
potential~\cite{Mes08b}. The extension to the time dependent case is
on the way.

Acknowledgments: This work was supported by the DFG, project nr. RE
322/10-1, the French-German exchange program PROCOPE nr. 07523TE, the
CNRS Programme ``Mat\'eriaux'' (CPR-ISMIR), Institut Universitaire de
France, the Humboldt foundation, a Gay-Lussac price, the French Agence
Nationale de la Recherche (ANR-06-BLAN-0319-02), and the French
computational facilities CalMip (Calcul en Midi-Pyr\'en\'ees), IDRIS
and CINES.

\newpage
\begin{appendix}

{

\section{Hermiticity of orthonormalization constraints}
\label{app:hermconstr}

The constraint on orthonormality of the single-particle
wavefunctions, or correspondingly unitarity of transformation
coefficients, introduces a matrix of Lagrangian multipliers and that
matrix ought to be hermitian. We prove that here for the case of the
time-dependent variational principle (\ref{eq:varprinconstr}).  Let us
split the matrix of Lagrange parameters $\lambda$ into hermitian part
$\mu$ and anti-hermitian part $\kappa$ as
$
\lambda_{\gamma\beta}
=
\mu_{\gamma\beta}
+
\kappa_{\gamma\beta}
$.
The variational principle thus becomes
$0=\delta S$ for the action
\begin{eqnarray}
S
=
\int_{t_0}^{t} \textrm dt' \Big(
 \sum_{\alpha}(\psi_\alpha|\mathrm{i}\hbar \partial_t|\psi_{\alpha})
  -
  E_\mathrm{SIC}
  +
  \sum_{\beta,\gamma}^{}(\psi_{\beta}|\psi_{\gamma})
  \big(\mu_{\gamma\beta}+\kappa_{\gamma\beta}\big)
 \Big)
 \quad.
\label{eq:varprincfull}
\end{eqnarray}
The action $S$ subsequently splits into real and imaginary part where
the latter reads simply
\begin{equation}
\Im\{S\}
=
\int_{t_0}^{t}\textrm dt' 
 \sum_{\beta,\gamma}^{}(\psi_{\beta}|\psi_{\gamma})
  \kappa_{\gamma\beta}
  \quad.
\end{equation}
Both parts are to be varied independently. Variation of the
imaginary part yields
%\begin{equation}
$$
  0
  =
  \delta_{\psi_\beta^*}\Im\{S\}
  \quad\Longrightarrow\quad
  \sum_\gamma\psi_{\gamma}(\mathbf{r})
  \kappa_{\gamma\beta}
  =
  0
  \quad \forall
  \beta,\mathbf{r}
  \quad\Longrightarrow\quad
  \kappa_{\gamma\beta}
  =
  0
  \quad.
$$
%\end{equation}
This means that we always have $\Im\{S\}=0$ and we deal
with a purely real action
\begin{eqnarray}
S
=
\int_{t_0}^{t}\textrm dt' \Big(
 \sum_{\alpha}(\psi_\alpha|\mathrm{i}\hbar \partial_t|\psi_{\alpha})
  -
  E_\mathrm{SIC}
  +
  \sum_{\beta,\gamma}^{}(\psi_{\beta}|\psi_{\gamma})
  \lambda_{\gamma\beta}
 \Big)
 \quad,\quad
 \lambda_{\beta\gamma}^*
 =
 \lambda_{\gamma\beta}^{\mbox{}}
 \quad.
\label{eq:varprincreal}
\end{eqnarray}
The same reasoning applies to all other form of action used for TDSIC
paper and to the energy functional used for stationary SIC.

}

\section{Alternative derivations of TDSIC}
\label{sec:vartdsic2}

{In this appendix, we will present two alternative derivations
of TDSIC.  It is gratifying to see that different derivations all lead
to the same result.  The alternative routes also shed some new light
on the intrinsic properties of TDSIC.  }

\subsection{The Goedecker method}
\label{sec:derivexplic}

First, we will use the Goedecker method of
variation~\cite{Goe97}.
One starts from the SIC energy $E_{\rm SIC}$ as given in
Eq. (\ref{eq:fsicen}) and defined in term of the the orthonormal set
of single particle wavefunctions $\{\psi_\alpha\}$.  Thanks to the
L\"owdin orthonormalization method~\cite{Low50}, one could equally
well expand the $\{\psi_\alpha\}$ into  a set of
non-orthogonal functions $\{\tilde{\psi}_\alpha\}$ as
\begin{eqnarray}
  \psi_\alpha=\sum_\beta s_{\beta,\alpha}^{-1/2}\tilde{\psi}_\beta
  \quad,\quad
  s_{\beta,\alpha}=(\tilde{\psi}_\beta|\tilde{\psi}_\alpha)
  \quad.
\end{eqnarray}
This is actually another way to constrain the orthonormality of the
$\{\psi_\alpha\}$.  
If one assumes that the $\{\tilde{\psi}_\alpha\}$ are not too
far from orthonormality, one has to first order 
\begin{eqnarray}
   \psi_\alpha
   \approx 
   \sum_\beta (\delta_{\alpha\beta}-\sigma_{\alpha\beta})\tilde{\psi}_\beta
   \quad,\quad
   \sigma_{\alpha\beta}
   =
   \frac{1}{2}\left(s_{\alpha\beta}-\delta_{\alpha\beta}\right)
   =
   \frac{1}{2}\left((\tilde{\psi}_\beta|\tilde{\psi}_\alpha)
      -\delta_{\alpha\beta}\right)
   \quad,
\label{eq:expand}
\end{eqnarray}
where $\sigma_{\alpha\beta}$ is a small quantity.
One now applies the principle of stationary action when varying with
respect to the $\tilde{\psi}^*_\alpha$ (the non-orthogonal
functions). This yields
\begin{eqnarray}
  0
  &=&
  \delta_{\tilde{\psi}_\alpha^*}
  \int_0^t \textrm dt'
    (\sum_\beta(\psi_\beta|\textrm i \hbar \partial_t\psi_\beta)
    -
    E_{\rm SIC})
  \quad.
\end{eqnarray}
Using the chain rule for functional derivatives, we obtain
\begin{equation}
  0
  =
  \sum_\gamma \int \textrm d^3 \mathbf{r'}
  \left( \frac{\delta\psi_\gamma(\mathbf{r'})}
       {\delta\tilde{\psi}^*_\alpha(\mathbf{r})} 
   \frac{\delta}{\delta\psi_\gamma(\mathbf{r'})} +
   \frac{\delta\psi^*_\gamma(\mathbf{r'})}
        {\delta\tilde{\psi}^*_\alpha(\mathbf{r})} 
   \frac{\delta}{\delta\psi^*_\gamma(\mathbf{r'})} \right)
  \left( \sum_\beta(\psi_\beta|\textrm i \hbar \partial_t\psi_\beta) 
    -E_{\rm SIC} \right)
  \quad.
\nonumber
\end{equation}
The variations with respect to the orthonormal set
$\delta\psi_\beta$ are similar to those used before in
the derivations of TDSIC. Its evaluation yields
\begin{equation*}
  0
  =
  \sum_\beta
  \int \textrm d^3 \mathbf r\left\{
  \frac{\delta\psi_\gamma^*(\mathbf{r})}
       {\delta\tilde{\psi}_\alpha^*(\mathbf{r}')}
  \left(\textrm i \hbar \partial_t-\hat{h}_\gamma\right)\psi_\gamma
  -
  \psi_\gamma^*\left(
   \textrm i \hbar\stackrel{\leftarrow}{\partial}_t+\hat{h}_\gamma\right)
  \frac{\delta\psi_\gamma(\mathbf{r})}
       {\delta\tilde{\psi}_\alpha^*(\mathbf{r}')}
  \right\}
  \quad.
\end{equation*}
The crucial step is now to evaluate the $\delta\tilde{\psi}_\alpha$
derivatives.  Using Eq. (\ref{eq:expand}) and taking into account
that $\delta\tilde{\psi}_\alpha^*$ appears not only explicitly in the
expansion, but also implicitly in the coefficients
$\sigma_{\alpha\beta}$, yields
\begin{eqnarray*}
  \frac{\delta\psi_\beta^*(\mathbf{r})}
       {\delta\tilde{\psi}_\alpha^*(\mathbf{r}')}
  &=&
  \delta^3(\mathbf{r}\!-\!\mathbf{r}')
  \left[
    \delta_{\alpha\beta}
    -\sigma_{\alpha\beta}
  \right]
   - 
   \frac{1}{2}
   \sum_k\psi_k(\mathbf{r}')\tilde{\psi}_k^*(\mathbf{r})
    \delta_{\alpha\beta}
  \;,
\\
  \frac{\delta\psi_\beta(\mathbf{r})}
       {\delta\tilde{\psi}_\alpha^*(\mathbf{r}')}
  &=&
  -
  \frac{1}{2}
  \tilde{\psi}_\alpha(\mathbf{r})\psi_\beta(\mathbf{r}')
  \;.
\end{eqnarray*}
Note that we can let $\sigma_{\alpha\beta}\longrightarrow 0$ after
variation. Thus we obtain finally the TDSIC equation
\begin{equation}
  0
  =
  (\mathrm{i} \hbar \partial_t-\hat{h}_\alpha)|\psi_\alpha)
  - 
  \sum_\beta |\psi_\beta)
  (\psi_\beta|\mathrm{i} \hbar \partial_t-
     \frac{\hat{h}_\alpha+\hat{h}_\beta}{2}|\psi_\alpha)
\label{eq:Goed1}
\end{equation}
The symmetry condition is recovered by projecting Eq. (\ref{eq:Goed1})
on $(\psi_\beta|$.

\subsection{Unitary variation}
\label{sec:altvar}

Section \ref{sec:derivexplic} uses a variation where orthonormality
is explicitly obeyed which allows to work without Lagrangian
parameters. There is an interesting alternative for such a
technique. One deals with a unitary variation according to Thouless
theorem \cite{Tho60}. Any variation from one Slater state
$|\Phi\rangle$ to another Slater state $|\Phi'\rangle$ can be
expressed as
\begin{subequations}
\begin{equation}
  |\Phi'\rangle
  =
  \exp{(\mathrm{i}\hat{A})}|\Phi\rangle
  \quad,\quad
  \hat{A}^\dagger
  =
  \hat{A}
  \quad.
\end{equation}
A variation is a small change and thus the varied states can be
obtained from linearization. This means
\begin{equation}
  |\delta\Phi\rangle
  =
  \mathrm{i}\hat{A}|\Phi\rangle
  \quad,\quad
  \langle\delta\Phi|
  =
  -\mathrm{i}\langle\Phi|\hat{A}^\dagger
  =
  -\mathrm{i}\langle\Phi|\hat{A}
  \quad,
\end{equation}
and subsequently in terms of single-particle wavefunctions
\begin{eqnarray}
  |\delta\psi_\alpha)
  &=&
  \mathrm{i}\sum_{n=1}^\infty|\delta\psi_n)A_{n\alpha}
  =
  \mathrm{i} \sum_{\beta\in\mathrm{occ}} |\delta\psi_\beta)A_{\beta\alpha}
  +
  \mathrm{i}\hat{A}_\perp|\delta\psi_\alpha)
  \quad,\quad
\\
  (\delta\psi_\alpha|
  &=&
  -\mathrm{i}\sum_{n=1}^\infty A_{\alpha n}(\delta\psi_n|
  =
  -\mathrm{i} \sum_{\beta\in\mathrm{occ}} A_{\alpha\beta}(\delta\psi_\beta|
  -
  \mathrm{i}(\delta\psi_\alpha|\hat{A}_\perp
  \quad,
\\
  \hat{A}_\perp
  &=&
  \hat{A}\hat{\Pi}_\perp
  +
  \hat{\Pi}_\perp\hat{A}
  \quad,
\end{eqnarray}
\end{subequations}
where one should be aware of the different ranges of
summations, the $n$ running over all single-particle space and
the $\beta$ only of occupied states.
The operator $\hat{A}_\perp$ is the part of the operator leading into
space orthogonal to the occupied states.  Variation then corresponds
to a free variation of the matrix elements of the hermitian operator
$\hat{A}$ (as long as hermiticity is obeyed). However, the limitation
to hermiticity means that $|\delta\psi_\alpha)$ and
$(\delta\psi_\alpha|$ cannot be varied independently anymore.

Applying that variation to the principle of stationary action yields
\begin{eqnarray*}
  0
  &=&
  \sum_{\beta\in\mathrm{occ}}\left\{
  -\mathrm{i}(\psi_\beta|\hat{A}
  \left(\mathrm{i} \hbar \partial_t-\hat{h}_\beta\right)|\psi_\beta)
  -\mathrm{i}
  (\psi_\beta|\left(
   \mathrm{i} \hbar \stackrel{\leftarrow}{\partial}_t+\hat{h}_\beta\right)
  \hat{A}|\psi_\beta)
  \right\}
\\
  &=&
  \sum_{\beta\in\mathrm{occ}}\sum_{n=1}^\infty\left\{
  -\mathrm{i}A_{\beta n}(\psi_n|
  \left(\mathrm{i} \hbar \partial_t-\hat{h}_\beta\right)|\psi_\beta)
  -\mathrm{i}
  (\psi_\beta|\left(
   \mathrm{i} \hbar \stackrel{\leftarrow}{\partial}_t+\hat{h}_\beta\right)
  |\psi_n)A_{n\beta}
  \right\}
\\
  &=&
  \sum_{\beta\in\mathrm{occ}}\sum_{n\perp\,\mathrm{occ}}\left\{
  -\mathrm{i}A_{\beta n}(\psi_n|
  \left(\mathrm{i} \hbar \partial_t-\hat{h}_\beta\right)|\psi_\beta)
  -\mathrm{i}
  (\psi_\beta|\left(
   \mathrm{i} \hbar \stackrel{\leftarrow}{\partial}_t+\hat{h}_\beta\right)
  |\psi_n)A_{n\beta}
  \right\}
\\
  &&
  -\mathrm{i}\sum_{\beta\in\mathrm{occ}}A_{\beta\alpha}(\psi_\alpha|
  \left(\mathrm{i} \hbar \partial_t-\hat{h}_\beta\right)|\psi_\beta)
  -
  \underbrace{
   \mathrm{i}\sum_{\beta\in\mathrm{occ}}
   (\psi_\beta|\left(
    \mathrm{i} \hbar \stackrel{\leftarrow}{\partial}_t+\hat{h}_\beta\right)
   |\psi_\alpha)A_{\alpha\beta}
  }_{
   \mathrm{i}\sum_{\beta\in\mathrm{occ}}
   (\psi_\alpha|\left(
    \mathrm{i} \hbar\partial_t+\hat{h}_\alpha\right)
   |\psi_\beta)A_{\beta\alpha}
  }
\\
  &=&
  \sum_{\beta\in\mathrm{occ}}\sum_{n\perp\,\mathrm{occ}}\left\{
  -\mathrm{i}A_{\beta n}(\psi_n|
  \left(\mathrm{i} \hbar \partial_t-\hat{h}_\beta\right)|\psi_\beta)
  -\mathrm{i}
  (\psi_\beta|\left(
   \mathrm{i} \hbar \stackrel{\leftarrow}{\partial}_t+\hat{h}_\beta\right)
  |\psi_n)A_{n\beta}
  \right\}
\\
  &&
  +\mathrm{i}\sum_{\beta\in\mathrm{occ}}A_{\beta\alpha}(\psi_\alpha|
            \hat{h}_\beta-\hat{h}_\alpha
  |\psi_\beta)
  \quad.
\end{eqnarray*}
Now, the matrix elements $A_{\beta n}$ and $A_{n\beta}$ can be varied
independently because $A_{n\beta}=A_{\beta n}^*\in\mathbb{C}$.
Similarly, all elements $A_{\beta\alpha}$ can be considered as being
independent. This yields the TDSIC equations as
\begin{subequations}
\label{eq:fullSICstep5}
\begin{eqnarray}
  \hat{\Pi}_\perp\mathrm{i} \hbar \partial_t\psi_\alpha
  &=&
  \hat{\Pi}_\perp\hat{h}_\alpha\psi_\alpha
  \quad,
\label{eq:fullSICstep5a}
\\
  0
  &=&
  K_{\alpha\beta}
  \quad.
\label{eq:fullSICstep5b}
\end{eqnarray}
\end{subequations}
{That, again, reproduces the TDSIC
equations (after proper rewriting) together with the symmetry
condition.}

\section{LDA exchange energy of the 1D model}
\label{sec:lda1d}

One has to evaluate the exchange-correlation energy as a functional of 
the local one-body density :
\begin{equation}
\label{eq:exc}
E_{\rm XC}[\rho]
=
\frac{1}{2} \int \textrm d^3\mathbf{r} \, \textrm d^3\mathbf{r}' 
\left(
\Gamma(\mathbf{r},\mathbf{r}')-\rho(\mathbf{r})
\rho(\mathbf{r}')
\right) 
\langle \mathbf{r}|\hat{v}|\mathbf{r}' \rangle
\end{equation}
For fermionic systems, in the general case : 
\begin{equation}
\Gamma(\mathbf{r},\mathbf{r}') = \langle
\psi|\hat{\Gamma}(\mathbf{r},\mathbf{r}')|\psi \rangle 
\end{equation}
where $\hat{\Gamma}(\mathbf{r},\mathbf{r}')=\sum_{i>j}^{} 
\{\delta(\mathbf r-\mathbf r_i)\delta(\mathbf r'-\mathbf r_j)+
\delta(\mathbf r-\mathbf r_j)\delta(\mathbf r'-\mathbf r_i)\}$ 
is the local two-body density matrix (which Pauli effects are included
in). In the following, we will focus on the exchange energy only, so
that we can limit the $\psi$ to Slater determinants.

In the 1D model, one now has to compute :
\begin{equation}
\Gamma(x,x') = \frac{1}{2} \sum_{i,j}^{} 
\langle ij|\hat{\Gamma}(x,x')|\widetilde{ij} \rangle,
\label{eq:gamma1D}
\end{equation}
with~:
\begin{equation}
\hat{\Gamma}(x,x')=\sum_{i>j}^{} \{\delta(x-x_i)\delta(x'-x_j)+
\delta(x-x_j)\delta(x'-x_i)\}
\end{equation}
For a 1D free gas, $\sum_{i}^{}$ becomes $\int \textrm dk_i
\frac{\gamma L}{2\pi}$ (where $L$ is the length of the box) and
$\displaystyle \langle ij|\hat{\Gamma}(x,x')|\widetilde{ij} \rangle =
\frac{2}{L^2} \left\{1 - \frac{1}{\gamma} \cos \left[(k_i-k_j)(x-x')
\right] \right\}$. Inserting this in Eq.~(\ref{eq:gamma1D}), one gets~:
\begin{equation}
\Gamma(x,x') = 
\left(\frac{\gamma}{2\pi} \right)^2
\int \textrm dk_i\, \textrm dk_j 
\left[ 1-\frac{1}{\gamma} \Re \left(e^{\textrm i(k_i-k_j)(x-x')}\right)
  \right]
\label{eq:gamma1D2}
\end{equation}
The first part of the integral gives ${k_F}^2$; the other part is
proportional to :
\begin{equation}
\int \textrm dk_i e^{\textrm i k_i(x-x')} 
\int \textrm dk_j e^{-\textrm i k_j(x-x')}
=\frac{e^{\textrm i k_F(x-x')}-1}{i(x-x')}
\times \frac{e^{-\textrm i k_F(x-x')}-1}{-\textrm i(x-x')}
=2\,\frac{1-\cos \left[k_F(x-x')\right]}{(x-x')^2}
\end{equation}
Collecting these results in (\ref{eq:gamma1D2}), one finds~:
\begin{equation}
\label{eq:gamma1D3}
\Gamma(x,x')=\rho_0^2 \left\{1-\frac{2}{\gamma}
D\left[k_F(x-x')\right] \right\} 
\end{equation}
with the 1D free gas density, $\rho_0=\frac{\gamma k_F}{2\pi}$, and
$D(x)=\frac{1-\cos x}{x^2}$.

We now insert (\ref{eq:gamma1D3}) in the 1D version of
(\ref{eq:exc}). If the 1D potential is a smoothed Coulomb one, such as
$\frac{1}{\sqrt{(x-x')^2+a}}$ (in Hartree units), a straightforward 
change of variable gives~:  
\begin{equation}
\rho\, \varepsilon_X(\rho)
= -\frac{\rho_0^2}{\gamma} \int_{-\infty}^{+\infty} \textrm dx 
\frac{D(k_F x)}{\sqrt{x^2+a}}
= -\frac{\gamma}{(2\pi)^2}\int_{-\infty}^{+\infty} \textrm dx 
\frac{1-\cos(\frac{2\pi\rho_0}{\gamma}x)}{x^2\sqrt{x^2+a}}.
\end{equation}
The LDA exchange potential then reads~:
\begin{equation}
U_{\rm LDA}^{\rm X} [\varrho]
= \frac{\delta}{\delta\rho}(\rho\, \varepsilon_X(\rho))|_{\rho=\varrho}
= -\frac{1}{2\pi}\int_{-\infty}^{+\infty} \textrm dx 
\frac{\sin(\frac{2\pi\varrho}{\gamma}x)}{x\sqrt{x^2+a}}
\end{equation}

\section{Proof of the conservation laws}
\label{sec:conserv_laws}

\subsection{Energy conservation}

The proof of energy conservation is straightforward
and does not need any comments~:
\begin{eqnarray*}
  \partial_t E_{\rm SIC}
  &=&
  \int \textrm d^3 \mathbf r\sum_i 
  \partial_t\varphi_i^*({\bf r},t)
  \frac{\delta E_{\rm SIC}}{\delta\varphi_i^*({\bf r},t)}
  +
  \mbox{c.c.}
\nonumber\\
  &=&
  \sum_i(\partial_t\varphi_i|\hat{h}_{\rm SIC}\varphi_i)
  +
  \sum_i(\hat{h}_{\rm SIC}\varphi_i|\partial_t\varphi_i)
\nonumber\\
  &=&
  -\mathrm i\sum_i(\hat{h}_{\rm SIC}\varphi_i|\hat{h}_{\rm SIC}\varphi_i)
  +
  \mathrm i\sum_i(\hat{h}_{\rm SIC}\varphi_i|\hat{h}_{\rm SIC}\varphi_i)
%\nonumber\\
%  &=&  
  =
  0
  \quad.
\end{eqnarray*}

\subsection{Orthonormality conservation}

The orthonormality should be conserved during time propagation because
we imposed it in the variation of the action. Nevertheless we will
check it explicitly.  Using the TDSIC resulting propagation scheme
(\ref{eq:csicprop-perturb}), we see that a sufficient condition for
the orthonormality to be conserved is that the propagator is unitary
within the space of occupied states, i.e.,
\begin{eqnarray}
(\varphi_i|e^{-\mathrm i \hat{h}_{\rm SIC}}|\varphi_j)
=
(\varphi_j|e^{-\mathrm i \hat{h}_{\rm SIC}}|\varphi_i)^* \quad .
\end{eqnarray}
This last equation is obviously verified in the occupied subspace (only), 
because of the "weak" hermiticity of $\hat{h}_{\rm SIC}$ in this
subspace as given by the symmetry condition
(\ref{eq:symcond2}). 

\subsection{Zero-Force Theorem}

The Zero-Force Theorem (ZFT) states that the kinetic energy plus the
electron-electron interaction part in the Kohn-Sham mean-field do not
change the total momentum of the electron cloud which is due to
translational symmetry of the electron-electron interaction and of the 
kinetic energy. We formulate that symbolically as
$\partial_t^\mathrm{(kin,el)} \langle \mathbf{p}\rangle =0$. 
The proof starts from the Kohn-Sham equations where all
external fields are dropped 
\begin{equation}
\left( \frac{\mathbf{p}^2}{2m}+\hat{U}^\mathrm{(el)} \right)
|\varphi_i)
= 
\mathrm i\hbar\partial_t^\mathrm{(kin,el)}|\varphi_i) \quad,
\end{equation}
and where $\hat{U}^\mathrm{(el)}$ is the Kohn-Sham mean-field
part stemming from the electrons.
The time change  of total momentum reads
\begin{eqnarray}
\partial_t^\mathrm{(kin,el)} \sum_i(\varphi_i|\mathbf{p}|\varphi_i)
&=&
\sum_i \left[
  (\partial_t^\mathrm{(kin,el)}\varphi_i|\mathbf{p}|\varphi_i) + 
  (\varphi_i|\mathbf{p}|\partial_t^\mathrm{(kin,el)}\varphi_i) \right]
\nonumber\\ 
&=&
\sum_i \left[
  (\partial_t^\mathrm{(kin,el)}\varphi_i|\mathbf{p}\varphi_i) + 
  (\mathbf{p}\varphi_i|\partial_t^\mathrm{(kin,el)}\varphi_i) \right]
\nonumber\\  
&=&
\frac{1}{\mathrm i\hbar} \sum_i \left[ -(U^\mathrm{(el)}
  \varphi_i|\mathbf{p}\varphi_i) 
  + (\mathbf{p}\varphi_i|U^\mathrm{(el)} \varphi_i)
  \right] 
\end{eqnarray}
The general form of the ZFT is thus, in $\{\mathbf{r}\}$ representation
\begin{eqnarray}
\sum_i \int \mathrm d^3 \mathbf{r} 
\left[
  (\mathbf{r}|U^\mathrm{(el)}|\varphi_i)^*\nabla\varphi_i(\mathbf{r})
  + 
  (\mathbf{r}|U^\mathrm{(el)}|\varphi_i)\nabla\varphi_i^*(\mathbf{r})
  \right] 
=0 
\label{eq:ZFT-gen0}
\end{eqnarray}

Now we check whether the TDSIC equation fulfills this theorem, where
$(\mathbf{r}|U^\mathrm{(el)}|\varphi_i)=\sum_\alpha u_{i\alpha}^*\psi_\alpha
U_{\rm LDA,el}[|\psi_\alpha|^2]$. After simple manipulations on
(\ref{eq:ZFT-gen0}), one obtains~: 
\begin{eqnarray}
\sum_i \int \textrm d^3 \mathbf{r} 
&& \left[
  (\mathbf{r}|U^\mathrm{(el)}|\varphi_i)^*\nabla\varphi_i(\mathbf{r})
  + 
  (\mathbf{r}|U^\mathrm{(el)}|\varphi_i)\nabla\varphi_i^*(\mathbf{r})
  \right] \nonumber\\ 
&=&
\sum_\alpha \int \textrm d^3\mathbf{r}
U_{\rm LDA,el}[|\psi_\alpha|^2]\nabla|\psi_\alpha|^2\nonumber\\
&=&
\sum_\alpha \int \textrm d^3 \mathbf{r} \, U_\alpha\nabla\rho_\alpha
\label{eq:ZFT-dsic2}
\end{eqnarray}
in compact notations.
However, since $U_\alpha$ is obtained variationally, we have~:
\begin{eqnarray}
\int \textrm d^3\mathbf{r} \, U_\alpha\nabla\rho_\alpha 
&=& 
\int \textrm d^3\mathbf{r} \, \textrm d^3 \mathbf{r'} 
\frac{\delta E}{\delta\rho_\alpha(\mathbf{r})} \frac{\delta
  \rho_\alpha(\mathbf{r})}{\delta \mathbf{r'}} \nonumber\\ 
&=& \int \textrm d^3 \mathbf{r'} 
\frac{\delta E}{\delta\mathbf{r'}} \nonumber\\
&=& 0
\end{eqnarray}
As a consequence, TDSIC does verify the ZFT relation
(\ref{eq:ZFT-gen0}), as any variational scheme should.

\end{appendix}

 \bibliographystyle{elsart-num}
%% %\bibliographystyle{unsrt}
 \bibliography{cluster,add}
%% % set a link to that site 

\end{document}